\documentclass[aps,floats]{revtex4}
\usepackage{amsmath,amssymb}
\usepackage{graphicx,epsfig}
\usepackage[greek,english]{babel}
\usepackage{bbold}

\begin{document}
\bibliographystyle {plain}

\pdfoutput=1
\def\oppropto{\mathop{\propto}} 
\def\opsimeq{\mathop{\simeq}}
\def\opoverderline{\mathop{\overline}}
\def\operarrow{\mathop{\longrightarrow}}
\def\opsim{\mathop{\sim}}

\def\fig#1#2{\includegraphics[height=#1]{#2}}
\def\figx#1#2{\includegraphics[width=#1]{#2}}


\title{  Inhomogeneous asymmetric exclusion processes between two reservoirs : \\ 
large deviations for the local empirical observables in the Mean-Field approximation

} 


\author{ C\'ecile Monthus }
 \affiliation{ Universit\'e Paris Saclay, CNRS, CEA, Institut de Physique Th\'eorique,
91191 Gif-sur-Yvette, France}

\begin{abstract}

For a given inhomogeneous exclusion processes on $N$ sites between two reservoirs, the trajectories probabilities allow to identify the relevant local empirical observables and to obtain the corresponding rate function at Level 2.5. In order to close the hierarchy of the empirical dynamics that appear in the stationarity constraints, we consider the simplest approximation, namely the Mean-Field approximation for the empirical density of two consecutive sites, in direct correspondence with the previously studied Mean-Field approximation for the steady state. For a given inhomogeneous Totally Asymmetric model (TASEP), this Mean-Field approximation yields the large deviations for the joint distribution of the empirical density profile and of the empirical current around the mean-field steady state; the further explicit contraction over the current allows to obtain the large deviations of the empirical density profile alone. For a given inhomogeneous Asymmetric model (ASEP), the local empirical observables also involve the empirical activities of the links and of the reservoirs; the further explicit contraction over these activities yields the large deviations for the joint distribution of the empirical density profile and of the empirical current. The consequences for the large deviations properties of time-additive space-local observables are also discussed in both cases.

\end{abstract}

\maketitle

\section{ Introduction  }

In the field of large deviations for the Markov dynamics of many-body models,
one should distinguish two different perspectives that can be summarized as follows.

(i) On one hand, one can consider the Markov process in the space of configurations
(for instance $2^N$ configurations for a system of $N$ classical spins)
and one can apply the explicit large deviations at Level 2.5 to characterize the joint distribution of the empirical density and of the empirical flows in the configuration space.
Indeed, while the initial classification involved only three levels 
(see the reviews \cite{oono,ellis,review_touchette} and references therein),
 with Level 1 for empirical observables,
Level 2 for the empirical density,
and Level 3 for the empirical process, 
the introduction of the Level 2.5 has been a major progress to characterize non-equilibrium steady states,
because the rate functions at Level 2.5 can be written explicitly for general Markov processes,
including discrete-time Markov chains
 \cite{fortelle_thesis,fortelle_chain,review_touchette,c_largedevdisorder,c_reset,c_inference},
continuous-time Markov jump processes
\cite{fortelle_thesis,fortelle_jump,maes_canonical,maes_onandbeyond,wynants_thesis,chetrite_formal,BFG1,BFG2,chetrite_HDR,c_ring,c_interactions,c_open,c_detailed,barato_periodic,chetrite_periodic,c_reset,c_inference,c_runandtumble,c_jumpdiff,c_skew,c_metastable,c_east}
and Diffusion processes 
\cite{wynants_thesis,maes_diffusion,chetrite_formal,engel,chetrite_HDR,c_reset,c_lyapunov,c_inference,c_skew,c_metastable,coghi}. From this explicit Level 2.5, many other large deviations properties can be derived
via the appropriate contraction. In particular, the Level 2 for the empirical density alone 
corresponds to the optimization of the Level 2.5 over the empirical flows.
More generally, the Level 2.5 allows to analyze the large deviations properties of any time-additive observable
via its decomposition in terms of the empirical density and flows.
The link with the alternative approach 
based on deformed Markov operators  \cite{derrida-lecture,sollich_review,lazarescu_companion,lazarescu_generic,jack_review,vivien_thesis,lecomte_chaotic,lecomte_thermo,lecomte_formalism,lecomte_glass,kristina1,kristina2,jack_ensemble,simon1,simon2,simon3,Gunter1,Gunter2,Gunter3,Gunter4,chetrite_canonical,chetrite_conditioned,chetrite_optimal,chetrite_HDR,touchette_circle,touchette_langevin,touchette_occ,touchette_occupation,derrida-conditioned,derrida-ring,bertin-conditioned,touchette-reflected,touchette-reflectedbis,c_lyapunov,previousquantum2.5doob,quantum2.5doob,quantum2.5dooblong,c_ruelle,lapolla}
 can be understood via the corresponding conditioned process obtained from the generalization of Doob's h-transform.

(ii) On the other hand, when the dynamical rules are local in space, 
one would like to analyze the dynamics via the $O(N)$ local densities and flows of the conserved quantities.
For instance, the Macroscopic Fluctuation Theory (see the review \cite{mft} and references therein)
is a renormalized theory for interacting driven diffusive systems in the hydrodynamic limit, where the 
action for dynamical trajectories is written as an integral over space and time 
of an elementary space-time local Lagrangian that is Gaussian with respect to the local current.
Similarly, for interacting random walkers in the continuous-time discrete-space framework, 
 space-time local Lagrangian have been analyzed in Refs \cite{c_interactions,chemical,chabane}.
 
In the present paper, we focus on asymmetric exclusion processes 
with space-dependent rates between two reservoirs
and we follow the approach that has recently been applied to the kinetically-constrained East model \cite{c_east} 
 in order to identify the appropriate local empirical observables
and to analyze whether it is possible to write closed large deviations properties for them.
Inhomogeneous exclusion processes have been already much studied,
either for samples with one or two bottlenecks \cite{1defect,greulich_defects,2defects},
or for samples with smoothly-varying hopping rates \cite{smooth,inhomo},
or for disordered samples  \cite{stanley,barma_short,barma_long,goldstein,krug,DTASEP,enaud,MFtasepHarris,rgshort,rglong,rgextreme,barma,MFtasep,greulich,MFasep,revisited,india,india2020,indian,bayesian}.
In particular, the Mean-Field approximation for the steady state has been applied to these various
inhomogeneous models and its validity has been tested via numerics
both for the Totally Asymmetric model (TASEP) in \cite{MFtasepHarris,MFtasep,indian,smooth,inhomo}
and for the Asymmetric model (ASEP) in \cite{MFasep,bayesian}.
Here we consider the analog Mean-Field approximation for the empirical dynamics
in order to obtain closed large deviations properties for local empirical observables.

The paper is organized as follows.
In section \ref{sec_models}, we introduce the notations for inhomogeneous exclusion processes on $N$ sites between two reservoirs and we recall the Mean-Field approximation at the level of the steady state.
In section \ref{sec_local}, we describe the properties of the relevant local empirical observables
that determine the trajectories probabilities and formulate the Mean-Field approximation
for the empirical density of two consecutive sites in order to obtain 
closed large deviations at Level 2.5  for the remaining local empirical observables. 
Section \ref{sec_asep} is devoted to the inhomogeneous Asymmetric model (ASEP), where
the Mean-Field rate function involves the 1-site empirical density, the local activities of the bonds,
and the global empirical current that flows through the whole sample.
In section \ref{sec_tasep}, we discuss the simplifications that occur
for the inhomogeneous Totally Asymmetric model (TASEP),
where the Mean-Field rate function involves only the 1-site empirical density and the global empirical current.
Our conclusions are summarized in section \ref{sec_conclusion}.
Appendix \ref{app_reminder2.5} contains a reminder on the large deviations at Level 2.5 for 
general continuous-time Markov jump processes.
The application to inhomogeneous exclusion processes between two reservoirs
is described in Appendix \ref{app_config} 
for the empirical observables defined in the whole configuration space,
in order to compare with the analysis of the main text based on local empirical observables.


\section{ Inhomogeneous exclusion processes on $N$ sites between two reservoirs   }

\label{sec_models}

\subsection{ Dynamical rates in the bulk and at the two boundaries }

Exclusion models involve particles with hard-core interaction.
It is convenient to denote the $2^N$ possible configurations via $N$ classical spins $S_i=\pm$ with $i=1,..,N$,
where $S_i=+$ represents a particle at position $i$, and $S_i=-$ represents a hole at position $i$.
The Markov generator of the continuous-time Markov Chain for these $N$ spins $(S_1,...,S_N)$
can be then written in terms of the local Pauli matrices associated to the $N$ spins
\begin{eqnarray}
W && = \sum_{i=1}^{N-1} \left[ w_{i+1/2}^+ \left( \sigma_i^- \sigma_{i+1}^+ -  \frac{(1+\sigma_i^z) (1-\sigma_{i+1}^z)} {4} \right)
+ w_{i+1/2}^- \left( \sigma_i^+ \sigma_{i+1}^- -  \frac{(1-\sigma_i^z) (1+\sigma_{i+1}^z)} {4} \right)
\right]
\nonumber \\
&& +  \left[ w_1^+ \left( \sigma_1^-  -  \frac{(1+\sigma_1^z) } {2} \right)
+ w_1^- \left( \sigma_1^+  -  \frac{(1-\sigma_1^z) } {2} \right) \right]
+ \left[ w_N^+ \left( \sigma_N^-  -  \frac{(1+\sigma_N^z) } {2} \right)
+ w_N^- \left( \sigma_N^+  -  \frac{(1-\sigma_N^z) } {2} \right) \right]
\label{markov}
\end{eqnarray}
with the following meaning for the space-dependent rates $w_.^{\pm}$.


\subsubsection{ Boundary dynamical rules : rates $w_1^{\pm}$ and $w_N^{\pm}$ on the boundary spins connected to the left and to the right reservoirs}

(L) The boundary spin $S_1$ in contact with the Left reservoir can flip from $S_1$ to $(-S_1)$ with rate $w_1^{S_1}$.
The physical meaning is that if the spin $S_1$ were isolated, the Left reservoir would impose the 
following probabilities $\pi_1^{S_1=\pm}$ to see $S_1=\pm$
\begin{eqnarray}
\pi_1^+ = \frac{ w_1^- }{w_1^+ + w_1^-} = 1-\pi_1^-
\label{leftproba}
\end{eqnarray}

(R) Similarly, the boundary spin $S_N$ in contact with the Right reservoir can flip from $S_N$ to $(-S_N)$ with rate $w_N^{S_N}$.
Again if the spin $S_N$ were isolated, the Right reservoir would impose the 
following probabilities $\pi_N^{S_N}$ to see $S_N=\pm$
\begin{eqnarray}
\pi_N^+ = \frac{ w_N^- }{w_N^+ + w_N^-} = 1-\pi_N^-
\label{rightproba}
\end{eqnarray}

\subsubsection{ Bulk dynamical rules : two rates $ w_{i+1/2}^{\pm} $ on each link $(i,i+1)$ for $i=1,..,N-1$ }

The bulk dynamics involves two rates $ w_{i+1/2}^{\pm} $ on each link $(i,i+1)$ for $i=1,..,N-1$ :
 the rate $w_{i+1/2}^{+} $ governs the flips of the pair $(S_i,S_{i+1}) $ from $(+-)$ to $(-+)$,
while the rate $w_{i+1/2}^{-} $ governs the flips of the pair $(S_i,S_{i+1}) $ from $(-+)$ to $(+-)$.


\subsection{ Special cases concerning the asymmetry of the bulk rates }

It is important to distinguish the following special cases of exclusion processes :

(a) the Symmetric model (SEP), where the two rates on each bulk link coincide
and can be interpreted as a local diffusion coefficient $ D_{i+1/2} $
\begin{eqnarray}
(SEP) \ \ \ \ \  w_{i+1/2}^{\pm} = D_{i+1/2}
\label{sep}
\end{eqnarray}
It can be maintained in a non-equilibrium state with a steady current
if the reservoirs rates tend to produce different densities on the two boundary sites (Eqs \ref{leftproba} and \ref{rightproba}). It should be stressed that this inhomogeneous Symmetric model is very special (with respect to the asymmetric models described below) because exact closed equations can be written for the dynamics of various observables like the density profile and the two-point correlations. So for the steady state in any given inhomogeneous sample, many explicit results
have been computed,
in particular the density profile, the two-point correlations, the mean current and the current fluctuations (see \cite{c_lindbladExclusion} and references therein).
As a consequence, the SEP case will not be considered in the present paper.

(b) the Asymmetric model (ASEP),
where the two bulk rates $ w_{i+1/2}^{\pm} $ on the link $(i+1/2)$ are different and non-vanishing,
can be better understood via the parametrization in terms of 
the local diffusion coefficient $D_{i+1/2} $ and the local force $ F_{i+1/2}$ on each link
\begin{eqnarray}
(ASEP) \ \ \ \ \  w_{i+1/2}^{\pm} && = D_{i+1/2} e^{ \pm \frac{F_{i+1/2} }{2 D_{i+1/2}} }
\label{asep}
\end{eqnarray}
The Mean-Field approximation for the steady state
has been studied for the random forces $F_{i+1/2} $ and unit diffusion coefficients $D_{i+1/2}=1$  in \cite{MFasep}
in relation with the Strong Disorder Renormalization Group (SDRG) studies \cite{rgshort,rglong,rgextreme},
while the Mean-Field approximation for the dynamics has been considered for biological applications 
in \cite{bayesian}.

(c) the Totally Asymmetric model (TASEP), where the rates corresponding to backward motion vanish, both for the bulk and for the boundary reservoirs
\begin{eqnarray}
(TASEP) \ \ \ \ \  w_{i+1/2}^{-} && = 0 \ \ {\rm for } \ \ i=1,..,N-1
 \nonumber \\
 w_1^+ && =0
  \nonumber \\
 w_N^- && =0
\label{tasep}
\end{eqnarray}
is fully irreversible and thus simpler than the Asymmetric model of Eq. \ref{asep}.
The Mean-Field approximation for the steady state has been analyzed for various inhomogeneous samples \cite{MFtasepHarris,MFtasep,indian,smooth,inhomo}.


\subsection{ Mean-Field approximation to obtain closed equations for the 1-spin density in the steady state   }

In order to better understand the physical meaning of the Mean-Field approximation at the level of empirical dynamics that will be described in the next sections, 
it is useful as comparison to recall here the Mean-Field approximation at the level of the steady state
  \cite{MFasep,MFtasepHarris,MFtasep,indian,bayesian,smooth,inhomo}.

For the Markov generator of Eq. \ref{markov}, the steady state $P(S_1,...,S_N)$
in the full configuration space of the $N$ spins satisfies
\begin{eqnarray}
&& 0  = \partial_t P(S_1,...,S_N)
 \label{steadyconfig}
\nonumber \\
&& =  \sum_{i=1 }^{N-1} 
 \left( \delta_{S_i,-} \delta_{S_{i+1},+} -  \delta_{S_i,+} \delta_{S_{i+1},-}  \right)
 \left[ w_{i+1/2}^{+}   P(..S_{i-1},+,-,S_{i+2}..)
 - w_{i+1/2}^{-}   P(..S_{i-1},-,+,S_{i+2}..)  \right]
\nonumber
 \\
 && + \left( \delta_{S_1,-}  -  \delta_{S_1,+}   \right)
 \left[  w_1^{+}  P(+,S_2,..)  - w_1^{-}  P(-,S_2,..)   \right]
 + \left( \delta_{S_N,-}  -  \delta_{S_N,+}   \right)
 \left[ w_N^{+}  P(..,S_{N-1},+) - w_N^{-}  P(..,S_{N-1},-)   \right]
 \nonumber
\end{eqnarray}
The basic local observables one is the most interested in are
the probabilities for two consecutive spins at positions $(i,i+1)$ in the steady state
\begin{eqnarray}
P_{i,i+1}^{S_i,S_{i+1}} \equiv  
 \left[ \prod_{n=1}^{i-1} \sum_{S_n=\pm}  \right] \left[ \prod_{p=i+2}^{N}\sum_{S_p=\pm}   \right]  P(S_1,...,S_N)
\label{P2steady}
\end{eqnarray}
and the probabilities for a single spin at position $i$ in the steady state
\begin{eqnarray}
P_i^{S_i} \equiv \left[ \prod_{n=1}^{i-1} \sum_{S_n=\pm}  \right] \left[ \prod_{p=i+1}^{N}\sum_{S_p=\pm}   \right]  P(S_1,...,S_N) = \sum_{S_{i+1}=\pm} P_{i,i+1}^{S_i,S_{i+1}}= \sum_{S_{i-1}=\pm} P_{i-1,i}^{S_{i-1},S_{i}}
\label{P1steady}
\end{eqnarray}
To obtain the steady-state equations for the probabilities $P_i^+$,
one needs to sum Eq. \ref{steadyconfig} over the other $(N-1)$ spins 
to obtain for the bulk spins $i=2,..,N-1$ 
\begin{eqnarray}
 0   =  \partial_t P_i^+ && = w_{i-1/2}^{+}   P_{i-1,i}^{+-}    - w_{i-1/2}^{-}   P_{i-1,i}^{-+}  
 - w_{i+1/2}^{+}   P_{i,i+1}^{+-}    + w_{i+1/2}^{-}   P_{i,i+1}^{-+}  
 \label{steadybulk}
\end{eqnarray}
 for the left boundary spins $i=1$ 
\begin{eqnarray}
 0   = \partial_t P_1^+ &&= w_1^{-} (1-P_1^+) - w_1^{+}  P_1^+
 - w_{3/2}^{+}   P_{1,2}^{+-}      +  w_{3/2}^{-}   P_{1,2}^{-+} 
 \label{steadyleft}
\end{eqnarray}
and for the right boundary spins $i=N$ 
\begin{eqnarray}
 0   = \partial_t P_N^+  = w_N^{-}  (1-P_{N}^+) - w_N^{+}  P_{N}^+
 + w_{N-1/2}^{+}   P_{N-1,N}^{+-}
 - w_{N-1/2}^{-}   P_{N-1,N}^{-+}
 \label{steadyright}
\end{eqnarray}

These stationarity equations for the 1-spin probabilities $P_i^+ $ involve the 2-spins probabilities $P_{..}^{..} $.
Similarly, the stationarity equations for the 2-spins probabilities $P_{..}^{..} $ will involve 3-spins probabilities,
so that one obtains a whole hierarchy.
In order to close this hierarchy, the simplest approximation is the Mean-Field approximation
where the 2-spins probabilities $P_{..}^{..} $ are approximated by the product of the 1-spin probabilities
\begin{eqnarray}
\left[ P_{i,i+1}^{S_i,S_{i+1}} \right]^{MF}  = P_i^{S_i} P_{i+1}^{S_i}
\label{P2steadyMF}
\end{eqnarray}
Plugging this Mean-Field approximation into Eqs \ref{steadybulk}
\ref{steadyleft} and \ref{steadyright}
leads to the following closed equations for the 1-spin probabilities $P_i^+=1-P_i^- $
\begin{eqnarray}
 0   =  \partial_t P_i^+ && = w_{i-1/2}^{+}   P_{i-1}^{+} (1-P_i^+)   - w_{i-1/2}^{-} (1-  P_{i-1}^{+}  )P_i^+
 - w_{i+1/2}^{+}   P_i^+(1-P_{i+1}^{+} )   + w_{i+1/2}^{-} (1-P_i^+)  P_{i+1}^{+}  
\nonumber \\
 0   = \partial_t P_1^+ &&= w_1^{-} (1-P_1^+) - w_1^{+}  P_1^+
 - w_{3/2}^{+}   P_1^+ (1-P_{2}^{+})     +  w_{3/2}^{-}   (1-P_1^+) P_{2}^{+} 
\nonumber \\
 0   = \partial_t P_N^+ && = w_N^{-}  (1-P_{N}^+) - w_N^{+}  P_{N}^+
 + w_{N-1/2}^{+}   P_{N-1}^{+} (1- P_{N}^{+})
 - w_{N-1/2}^{-} (1- P_{N-1}^{+})  P_{N}^{+} 
 \label{MFsteady}
\end{eqnarray}
that have been previously much studied both for ASEP \cite{MFasep,bayesian} and for TASEP \cite{MFtasepHarris,MFtasep,indian,smooth,inhomo}, in particular to test the validity
of this Mean-Field approximation against Monte-Carlo numerics
(while in pure exclusion models, the validity of the Mean-Field approximation 
has been discussed \cite{MFpure} via the comparison with exact solutions).


The goal of the present paper is to analyze the possible dynamical fluctuations
around the mean-field steady state of Eq. \ref{MFsteady}
in a given inhomogeneous sample defined by the space-dependent rates $w_.^.$.


\section{ Analysis of the relevant local empirical time-averaged observables }

\label{sec_local}

In this section, we follow the approach that has recently been applied to the kinetically-constrained East model \cite{c_east} in order to identify the appropriate local empirical observables
and to analyze whether it is possible to write closed large deviations properties for them.

\subsection{ Identification of the relevant time-empirical observables that determine the trajectories probabilities }

For a given Markov model, the relevant time-empirical observables
are defined as the time-empirical observables that determine the trajectories probabilities (see Appendix A of
\cite{c_east} for a general discussion).
For the present Markov jump process,  
the probability of Eq. \ref{pwtrajjump} for the trajectory 
$C(t)=\{S_1(t),..,S_i(t),.,S_N(t) \}$ during the time-window $0 \leq t \leq T$
reads
\begin{eqnarray}
\ln \left( {\cal P}[C(0 \leq t \leq T)]   \right)
&& =     \sum_{t \in [0,T] : S_1(t^+) \ne S_1(t) }   \ln \left( w_1^{S_1(t)}  \right)
+  \sum_{t \in [0,T] : S_N(t^+) \ne S_N(t) }   \ln \left( w_N^{S_N(t)}  \right)
\nonumber \\
&& +  \sum_{i=1}^{N-1} \sum_{t \in [0,T] : \substack{ S_i(t)=+ ;  S_{i+1}(t)=-   \\ S_i(t^+)= - ; S_{i+1}(t^+)=+}}    \ln \left( w_{i+1/2}^+  \right)
+ \sum_{i=1}^{N-1} \sum_{t \in [0,T] : \substack{ S_i(t)=- ;  S_{i+1}(t)=+   \\ S_i(t^+)= + ; S_{i+1}(t^+)=-}}    \ln \left( w_{i+1/2}^-  \right)
\nonumber \\
&&  -   \int_0^T dt \left[ w_1^{S_1(t)} +w_N^{S_N(t)} 
+\sum_{i=1}^{N-1} \left( w_{i+1/2}^+ \delta_{S_i(t),+} \delta_{S_{i+1}(t),-} 
+ w_{i+1/2}^- \delta_{S_i(t),-} \delta_{S_{i+1}(t),+}  \right) \right]   
\label{pwtrajspin}
\end{eqnarray}
As a consequence, this probability can be rewritten as
\begin{eqnarray}
 {\cal P}[C(0 \leq t \leq T)] \opsimeq_{T \to +\infty}   
 e^{\displaystyle  -T  \Phi_{[w_.^.]}  \left( \rho_.^. ; \rho_{..}^{..} ;q_.^.  \right) }
\label{ptrajectempi}
\end{eqnarray}
where the action 
\begin{eqnarray}
&& \Phi_{[w_.^.]}  \left( \rho_.^. ; \rho_{..}^{..} ;q_.^.  \right) 
 =  
    \left[ w_1^+ \rho_1^+ + w_1^- \rho_1^- + w_N^+ \rho_N^+ + w_N^- \rho_N^-
+\sum_{i=1}^{N-1} \left( w_{i+1/2}^+  \rho_{i,i+1}^{+-} +w_{i+1/2}^-  \rho_{i,i+1}^{-+}   \right) \right]   
\nonumber \\
&&  
 -  q_1^+   \ln \left( w_1^+  \right) -  q_1^-   \ln \left( w_1^-  \right)
-  q_N^+   \ln \left( w_N^+  \right) -  q_N^-   \ln \left( w_N^-  \right) 
-   \sum_{i=1}^{N-1}  \left[ q_{i+1/2}^+  \ln \left( w_{i+1/2}^+  \right) + q_{i+1/2}^-  \ln \left( w_{i+1/2}^-  \right)
 \right]
\label{pwtrajspinempi}
\end{eqnarray}
contains the rates $w_.^. $ as parameters
and the following local empirical time-averaged observables $\left( \rho_.^. ; \rho_{..}^{..} ;q_.^.  \right) $ as variables:

(i) the empirical time-averaged densities for the single spin $S_i$ at position $i$
\begin{eqnarray}
\rho_{i}^{S_i} \equiv \frac{1}{ T }    \int_0^T dt \ \delta_{S_{i}(t) ,S_i} 
\label{rho1def}
\end{eqnarray}
and for two consecutive spins $(S_i,S_{i+1})$ at positions $(i,i+1)$
\begin{eqnarray}
\rho_{i,i+1}^{S_i,S_{i+1}} \equiv \frac{1}{ T }    \int_0^T dt 
\ \delta_{S_{i}(t) ,S_i} \ \delta_{S_{i+1}(t) ,S_{i+1}} 
\label{rho2def}
\end{eqnarray}

(ii) the local empirical time-averaged flows associated to the bulk link $(i+1/2)$
\begin{eqnarray}
q_{i+1/2}^+  =  \frac{1}{ T }     \sum_{t \in [0,T] : \substack{ S_i(t)=+ ;  S_{i+1}(t)=-   \\ S_i(t^+)= - ; S_{i+1}(t^+)=+}}  1
\nonumber \\
q_{i+1/2}^-  =  \frac{1}{ T }     \sum_{t \in [0,T] : \substack{ S_i(t)=- ;  S_{i+1}(t)=+   \\ S_i(t^+)= + ; S_{i+1}(t^+)=-}}  1
\label{qbulklocal}
\end{eqnarray}
associated to the spin $S_1$ (left reservoir)
\begin{eqnarray}
q_1^{+}  \equiv  \frac{1}{T}     \sum_{t \in [0,T] :  \substack{S_1(t)=+ \\   S_1(t^+)= - }} 1
\nonumber \\
q_1^{-}  \equiv  \frac{1}{T}    \sum_{t \in [0,T] :  \substack{S_1(t)=- \\   S_1(t^+)= + }} 1
\label{qleftlocal}
\end{eqnarray}
and associated to the spin $S_N$ (right reservoir) 
\begin{eqnarray}
q_N^{+}  \equiv  \frac{1}{T}     \sum_{t \in [0,T] :  \substack{S_N(t)=+ \\  S_N(t^+)= - }} 1
\nonumber \\
q_N^{-}  \equiv  \frac{1}{T}     \sum_{t \in [0,T] :  \substack{S_N(t)=- \\  S_N(t^+)= + }} 1
\label{qrightlocal}
\end{eqnarray}

In conclusion, the relevant empirical observables that determine the trajectories probabilities of Eq. \ref{pwtrajspin} are the local empirical densities of Eqs \ref{rho1def} and \ref{rho2def}
and the local empirical flows of Eqs \ref{qbulklocal} \ref{qleftlocal} and \ref{qrightlocal},
that only involves one spin or two consecutive spins.


\subsection{ Typical values of the relevant time-empirical observables $\left( \rho_.^. ; \rho_{..}^{..} ;q_.^.  \right) $ for the rates $[w_.^.] $ }

The typical values of the local empirical densities $\left( \rho_.^. ; \rho_{..}^{..}  \right)$ of Eqs \ref{rho1def}
and \ref{rho2def}
are given by the one-spin probabilities $P_.^. $ and the two-spins probabilities $P_{..}^{..} $ in the steady state  
discussed in Eqs \ref{P2steady} and \ref{P1steady}
\begin{eqnarray}
 \left[\rho_{i}^{S_i} \right]^{typ} &&= P_i^{S_i} 
\nonumber \\
 \left[ \rho_{i,i+1}^{S_i,S_{i+1}}  \right]^{typ} &&=  P_{i,i+1}^{S_i,S_{i+1}} 
\label{typlocalrho}
\end{eqnarray}

The typical values of the local empirical flows $q_.^.$ of Eqs \ref{qbulklocal} \ref{qleftlocal} and \ref{qrightlocal}
are given by the 
steady state flows $Q_.^.(...) $ that can be evaluated from the rates $w_.^.$ and from the steady state 
probabilities $P_.^. $ and $P_{..}^{..} $
\begin{eqnarray}
 \left[q_{i+1/2}^{+} \right]^{typ} && = Q_{i+1/2}^{+}    \equiv  w_{i+1/2}^{+}   P_{i,i+1}^{+-}
\nonumber \\
 \left[ q_{i+1/2}^{-} \right]^{typ} && = Q_{i+1/2}^{-}   \equiv  w_{i+1/2}^{-}   P_{i,i+1}^{-+}
\nonumber \\
 \left[q_1^{S_1} \right]^{typ} && = Q_1^{S_1}  \equiv  w_1^{S_1}  P_1^{S_1}
\nonumber \\
 \left[ q_N^{S_N}\right]^{typ} && = Q_N^{S_N}   \equiv  w_N^{S_N}  P_N^{S_N} 
\label{typlocalq}
\end{eqnarray}


\subsection{ Number of dynamical trajectories of length $T$ with the same local empirical observables $\left( \rho_.^. ; \rho_{..}^{..} ;q_.^.  \right) $  }

Since all the individual dynamical trajectories $ C(0 \leq t \leq T)  $
 that have the same local empirical observables $\left( \rho_.^. ; \rho_{..}^{..} ;q_.^.  \right) $ 
  have the same probability given by Eq. \ref{ptrajectempi},
 one can rewrite the normalization over all possible trajectories 
as a sum over these empirical observables
\begin{eqnarray}
1= \sum_{C(0 \leq t \leq T)}  {\cal P}[C(0 \leq t \leq T)] 
\opsimeq_{T \to +\infty}  
 \sum_{\left( \rho_.^. ; \rho_{..}^{..} ;q_.^.  \right)} \Omega_T\left( \rho_.^. ; \rho_{..}^{..} ;q_.^.  \right) 
 e^{\displaystyle  -T  \Phi_{[w_.^.]}  \left( \rho_.^. ; \rho_{..}^{..} ;q_.^.  \right) }
\label{normaempi}
\end{eqnarray}
where the number $\Omega_T\left( \rho_.^. ; \rho_{..}^{..} ;q_.^.  \right)  $ of dynamical trajectories of length $T$ associated to given values $\left( \rho_.^. ; \rho_{..}^{..} ;q_.^.  \right) $ of these 
empirical observables
 is expected to grow exponentially with respect to the length $T$ of the trajectories
\begin{eqnarray}
 \Omega_T\left( \rho_.^. ; \rho_{..}^{..} ;q_.^.  \right) \opsimeq_{T \to +\infty} C\left( \rho_.^. ; \rho_{..}^{..} ;q_.^.  \right) \ e^{\displaystyle T S\left( \rho_.^. ; \rho_{..}^{..} ;q_.^.  \right)  }
\label{omegat}
\end{eqnarray}
The prefactor $C\left( \rho_.^. ; \rho_{..}^{..} ;q_.^.  \right)$ denotes the appropriate constitutive constraints for the empirical observables $\left( \rho_.^. ; \rho_{..}^{..} ;q_.^.  \right)$ that will be discussed later.
The factor $S\left( \rho_.^. ; \rho_{..}^{..} ;q_.^.  \right)  = \frac{\ln \Omega_T\left( \rho_.^. ; \rho_{..}^{..} ;q_.^.  \right) }{ T }  $ represents the 
Boltzmann intensive entropy of the set of trajectories of length $T$ with given empirical observables $\left( \rho_.^. ; \rho_{..}^{..} ;q_.^.  \right) $.
Let us now recall how it can be evaluated without any actual computation (i.e. one does not need 
to use combinatorial methods to count the appropriate configurations).

The normalization of Eq. \ref{normaempi} becomes for large $T$
\begin{eqnarray}
1  \opsimeq_{T \to +\infty} \sum_{\left( \rho_.^. ; \rho_{..}^{..} ;q_.^.  \right)} 
C\left( \rho_.^. ; \rho_{..}^{..} ;q_.^.  \right) \  e^{\displaystyle  T \left[ S\left( \rho_.^. ; \rho_{..}^{..} ;q_.^.  \right) - \Phi_{[w_.^.]}  \left( \rho_.^. ; \rho_{..}^{..} ;q_.^.  \right)  \right] }
\label{normaempit}
\end{eqnarray}
When the empirical variables $\left( \rho_.^. ; \rho_{..}^{..} ;q_.^.  \right) $ take their typical values 
$\left[ \rho_.^. ; \rho_{..}^{..} ;q_.^.  \right]^{typ}$ given by Eqs \ref{typlocalrho} and \ref{typlocalq},
the exponential behavior in $T$ of Eq. \ref{normaempit}
should exactly vanish,
i.e. the entropy $S \left( P_.^. ; P_{..}^{..} ;Q_.^.  \right) $ associated to these typical values
should exactly compensate the corresponding action $\Phi_{[w_.^.]} \left( P_.^. ; P_{..}^{..} ;Q_.^.  \right)   $ 
\begin{eqnarray}
S\left( P_.^. ; P_{..}^{..} ;Q_.^.  \right)= \Phi_{[w_.^.]} \left( P_.^. ; P_{..}^{..} ;Q_.^.  \right)   
\label{compensation}
\end{eqnarray}
To obtain the intensive entropy $S\left( \rho_.^. ; \rho_{..}^{..} ;q_.^.  \right)  $ 
for any other given values $\left( \rho_.^. ; \rho_{..}^{..} ;q_.^.  \right)   $ of the empirical observables,
one just needs to introduce the modified Markov rates $\hat w_.^.$ that would make 
the empirical values $\left( \rho_.^. ; \rho_{..}^{..} ;q_.^.  \right) = \left( \hat P_.^. ; \hat P_{..}^{..} ; \hat Q_.^.  \right)$ typical for this modified model,
i.e. the modified rates $\hat w_.^.$ can be evaluated from Eq. \ref{typlocalq}
\begin{eqnarray}
\hat  w_{i+1/2}^{+} && = \frac{ \hat Q_{i+1/2}^{+}  }{\hat P_{i,i+1}^{+-} } =  \frac{q_{i+1/2}^{+}  }{\rho_{i,i+1}^{+-} }
\nonumber \\
\hat w_{i+1/2}^{-} && = \frac{ \hat Q_{i+1/2}^{-}}{\hat P_{i,i+1}^{-+} } = \frac{ q_{i+1/2}^{-}}{\rho_{i,i+1}^{-+} }
\nonumber \\
\hat w_1^{S_1} && = \frac{ \hat Q_1^{S_1}  }{\hat P_1^{S_1} } = \frac{ q_1^{S_1}  }{\rho_1^{S_1} }
\nonumber \\
\hat w_N^{S_N} && = \frac{ \hat Q_N^{S_N}}{ \hat P_N^{S_N} }= \frac{ q_N^{S_N}}{ \rho_N^{S_N} }
\label{modeleeff}
\end{eqnarray}
Another interesting interpretation is that these modified rates $\hat w_.^.$
are the rates that would be inferred as the most probable model from the data $\left( \rho_.^. ; \rho_{..}^{..} ;q_.^.  \right) $ concerning the local empirical observables \cite{c_inference}.

We may now use Eq. \ref{compensation} for this modified model to obtain
\begin{eqnarray}
S\left( \rho_.^. ; \rho_{..}^{..} ;q_.^.  \right)
&& = S \left( \hat P_.^. ; \hat P_{..}^{..} ; \hat Q_.^.  \right)
=    \Phi_{[{\hat w}_.^.]}  \left( \hat P_.^. ; \hat P_{..}^{..} ; \hat Q_.^.  \right) 
=  \Phi_{[{\hat w}_.^.]} \left( \rho_.^. ; \rho_{..}^{..} ;q_.^.  \right) 
\label{entropyempi}
\end{eqnarray}
With the explicit form of Eq. \ref{pwtrajspinempi}
 for the action $\Phi_{[{\hat w}_.^.]} \left( \rho_.^. ; \rho_{..}^{..} ;q_.^.  \right) $,
 where we can plug the explicit modified rates $\hat w_.^.$ of Eq. \ref{modeleeff},
 one obtains the entropy $S\left( \rho_.^. ; \rho_{..}^{..} ;q_.^.  \right) $ 
 as a function of the local empirical observables $\left( \rho_.^. ; \rho_{..}^{..} ;q_.^.  \right) $
\begin{eqnarray}
&& S\left( \rho_.^. ; \rho_{..}^{..} ;q_.^.  \right)  = \Phi_{[\hat w_.^.]}  \left( \rho_.^. ; \rho_{..}^{..} ;q_.^.  \right) 
\nonumber \\
&&  =  
    \left[ \hat w_1^+ \rho_1^+ + \hat w_1^- \rho_1^- + \hat w_N^+ \rho_N^+ + \hat w_N^- \rho_N^-
+\sum_{i=1}^{N-1} \left( \hat w_{i+1/2}^+  \rho_{i,i+1}^{+-} +\hat w_{i+1/2}^-  \rho_{i,i+1}^{-+}   \right) \right]   
\nonumber \\
&&  
 -  q_1^+   \ln \left( \hat w_1^+  \right) -  q_1^-   \ln \left( \hat w_1^-  \right)
-  q_N^+   \ln \left( \hat w_N^+  \right) -  q_N^-   \ln \left( \hat w_N^-  \right) 
-   \sum_{i=1}^{N-1}  \left[ q_{i+1/2}^+  \ln \left( \hat w_{i+1/2}^+  \right) - q_{i+1/2}^-  \ln \left( \hat w_{i+1/2}^-  \right)
 \right]
 \nonumber \\
 && =   \left[ q_1^+ + q_1^- + q_N^+ + q_N^-
+\sum_{i=1}^{N-1} \left( q_{i+1/2}^+   +q_{i+1/2}^-     \right) \right]   
\nonumber \\
&&  
 -  q_1^+   \ln \left( \frac{ q_1^+  }{\rho_1^+ }  \right) -  q_1^-   \ln \left( \frac{ q_1^-  }{\rho_1^- }  \right)
-  q_N^+   \ln \left( \frac{ q_N^+  }{\rho_N^+ }  \right) -  q_N^-   \ln \left( \frac{ q_N^-  }{\rho_N^- }   \right) 
-   \sum_{i=1}^{N-1}  \left[ q_{i+1/2}^+  \ln \left(  \frac{q_{i+1/2}^{+}  }{\rho_{i,i+1}^{+-} }  \right)
 + q_{i+1/2}^-  \ln \left(  \frac{q_{i+1/2}^{-}  }{\rho_{i,i+1}^{-+} }  \right)
 \right] \ \ 
\label{entropyempiexplicit}
\end{eqnarray}


\subsection{ Rate function at Level 2.5 for the relevant local empirical observables $\left( \rho_.^. ; \rho_{..}^{..} ;q_.^.  \right) $  }

The normalization over trajectories of Eq \ref{normaempi} can be rewritten as the normalization
\begin{eqnarray}
 1  = \sum_{ \left( \rho_.^. ; \rho_{..}^{..} ;q_.^.  \right) }  P_T^{[2.5]}  \left( \rho_.^. ; \rho_{..}^{..} ;q_.^.  \right)
\label{normaprobaempi}
\end{eqnarray}
for the probability to see the local empirical observables $\left( \rho_.^. ; \rho_{..}^{..} ;q_.^.  \right)$
\begin{eqnarray}
P_T^{[2.5]} \left( \rho_.^. ; \rho_{..}^{..} ;q_.^.  \right) 
&& \opsimeq_{T \to +\infty}  
\Omega_T\left( \rho_.^. ; \rho_{..}^{..} ;q_.^.  \right) e^{\displaystyle  -T  \Phi_{[w_.^.]} \left( \rho_.^. ; \rho_{..}^{..} ;q_.^.  \right) }
\nonumber \\
&& \opsimeq_{T \to +\infty}  C\left( \rho_.^. ; \rho_{..}^{..} ;q_.^.  \right)  \ e^{\displaystyle  - T  I_{2.5} \left( \rho_.^. ; \rho_{..}^{..} ;q_.^.  \right)     }
\label{probempi2.5}
\end{eqnarray}
where the rate function at Level 2.5 the local empirical observables $\left( \rho_.^. ; \rho_{..}^{..} ;q_.^.  \right)$
reads using the explicit forms of Eqs \ref{pwtrajspinempi}
and \ref{entropyempiexplicit}
\begin{eqnarray}
&& I_{2.5} \left( \rho_.^. ; \rho_{..}^{..} ;q_.^.  \right)  
= \Phi_{[w_.^.]} \left( \rho_.^. ; \rho_{..}^{..} ;q_.^.  \right)  - S \left( \rho_.^. ; \rho_{..}^{..} ;q_.^.  \right)  
\nonumber \\
&& =  
 \sum_{i=1}^{N-1}  \left[ q_{i+1/2}^+  \ln \left(  \frac{q_{i+1/2}^{+}  }{w_{i+1/2}^+\rho_{i,i+1}^{+-} }  \right)
- q_{i+1/2}^+ + w_{i+1/2}^+  \rho_{i,i+1}^{+-} \right]
  +   \sum_{i=1}^{N-1}  \left[ q_{i+1/2}^-  \ln \left(  \frac{q_{i+1/2}^{-}  }{w_{i+1/2}^-\rho_{i,i+1}^{-+} }  \right)
- q_{i+1/2}^- + +w_{i+1/2}^-  \rho_{i,i+1}^{-+} \right]
\nonumber \\
&& +
  \left[  q_1^+   \ln \left( \frac{ q_1^+  }{ w_1^+\rho_1^+ }  \right) - q_1^+ + w_1^+ \rho_1^+ \right]
 +  \left[ q_1^-   \ln \left( \frac{ q_1^-  }{ w_1^-  \rho_1^- }  \right) - q_1^- + w_1^- \rho_1^- \right]
\nonumber \\
&& +   \left[ q_N^+   \ln \left( \frac{ q_N^+  }{ w_N^+ \rho_N^+ }  \right) - q_N^+ + w_N^+ \rho_N^+ \right]
+   \left[ q_N^-   \ln \left( \frac{ q_N^-  }{w_N^-\rho_N^- }   \right) - q_N^- + w_N^- \rho_N^- \right]
  \label{rateempi}
\end{eqnarray}
For each rate $w_.^.$ of the model, one recognizes 
the standard relative entropy cost of having a corresponding empirical flow $q_.^. $
different from the typical flow $w_.^.  \rho_{..}^{..} $ or $w_.^.  \rho_.^. $
 that would be produced by the local empirical densities.
 
In the two next subsections, we need to analyze the constitutive constraints $C\left( \rho_.^. ; \rho_{..}^{..} ;q_.^.  \right) $
that appear in the large deviations of Eq. \ref{probempi2.5}.


\subsection{ Closed constitutive constraints for the empirical 1-spin density $\rho_.^.$ and 2-spin density $\rho_{..}^{..} $}

The empirical 1-spin density $\rho_{i}^{S_i} $ of Eq. \ref{rho1def}
satisfies the normalization (Eq. \ref{rho1norma})
\begin{eqnarray}
\sum_{S_i=\pm} \rho_{i}^{S_i} = \rho_{i}^+ + \rho_i^- =1
\label{rho1norma1}
\end{eqnarray}

The empirical 2-spin density 
$\rho_{i,i+1}^{S_i,S_{i+1}} $ of Eq. \ref{rho2def} 
 should be compatible with the 1-spin empirical density of Eq. \ref{rho1def} via the summation over one spin
\begin{eqnarray}
\sum_{S_{i+1}=\pm} \rho_{i,i+1}^{S_i,S_{i+1}} && =\rho_{i}^{S_i} 
\nonumber \\
\sum_{S_i=\pm} \rho_{i,i+1}^{S_i,S_{i+1}} && =\rho_{i+1}^{S_{i+1}} 
\label{rho1from2}
\end{eqnarray}


\subsection{ Closure problem in the stationary constraints for the local empirical flows $q_.^.$}

\label{subsec_closed1}

It is now convenient to use the standard parametrization of
 flows in terms of activities and currents,
since only the current contributions are involved in stationarity constraints.

\subsubsection{ Parametrization of the local empirical flows $q_.^.$ in terms of the empirical activities $a_.$ and currents $j_.$ }

For each bulk link $(i+1/2)$ with $i=1,..,N-1$,  
the two empirical flows $q_{i+1/2}^+$ and $q_{i+1/2}^- $ can be parametrized
\begin{eqnarray}
 q_{i +1/2}^+  && =  \frac{   a_{i+1/2} +  j_{i+1/2}}{2} 
\nonumber \\
  q_{i-1/2}^+ && =  \frac{   a_{i+1/2}-j_{i+1/2}}{2}  
 \label{qaj}
\end{eqnarray}
 by their symmetric and antisymmetric parts called the empirical activity and the empirical current
\begin{eqnarray}
  a_{i+1/2}  \equiv   q_{i+1/2}^+ +   q_{i+1/2}^-  
  \nonumber \\
  j_{i+1/2}  \equiv   q_{i+1/2}^+ -   q_{i+1/2}^-  
\label{aj}
\end{eqnarray}
Similarly, the two empirical flows $q_1^{\pm}$ connected to the left reservoir
and the two empirical flows $q_N^{\pm}$ connected to the right reservoir
can be parametrized 
\begin{eqnarray}
  q_1^+ && =  \frac{   a_1-j_1}{2}  
  \nonumber \\
  q_1^- && =  \frac{   a_1+ j_1}{2}  
    \nonumber \\
  q_N^+ && =  \frac{   a_N+j_N}{2}  
  \nonumber \\
  q_N^- && =  \frac{   a_N -  j_N}{2}  
 \label{qajLR}
\end{eqnarray}
by the activities and the currents
\begin{eqnarray}
  a_1  \equiv   q_1^+ +   q_1^-  
  \nonumber \\
  j_1  \equiv   - q_1^+ +   q_1^-  
  \nonumber \\
   a_N  \equiv   q_N^+ +   q_N^-  
  \nonumber \\
  j_N  \equiv    q_N^+ -   q_N^-   
\label{ajbords}
\end{eqnarray}

\subsubsection{ Constraints on the local empirical currents $j_.$ to ensure the stationarity of the 1-spin empirical density $\rho_.^.$ }

The stationarity constraint of Eq. \ref{contrainteqs} for the empirical density $\rho(S_1,...,S_N)$
in the full configuration space is the empirical analog of the steady state Equation \ref{steadyconfig}.
The main difference is that the empirical flows are now independent variables with respect to the empirical densities.
Eq. \ref{contrainteqs} can be summed 
over $(N-1)$ spins to obtain the stationarity constraint for the 1-spin density $\rho_.^.$ as follows.

(i) For $ 2\leq i \leq N-1$,  the summation of Eq. \ref{contrainteqs} over the $(N-1)$ spins $(S_1,...,S_{i-1})$ and $(S_{i+1},..,S_N)$
yields the stationarity constraint for the 1-spin density $\rho_i^{+}=1-\rho_i^-$ 
\begin{eqnarray}
 0  =   \partial_t \rho_i^+ =  (  q_{i-1/2}^{+}   - q_{i-1/2}^{-}    ) 
 - (q_{i+1/2}^{+}   - q_{i+1/2}^{-} )  
 =   j_{i-1/2}  - j_{i+1/2}    
 \label{statiobulk}
\end{eqnarray}
This is the empirical analog of Eq. \ref{steadybulk} concerning the steady state.

(ii) The summation of Eq. \ref{contrainteqs} over the $(N-1)$ spins $(S_2,S_3,...,S_N)$ 
yields the stationarity constraint for the 1-spin density $\rho_1^{+}=1-\rho_1^-$ 
\begin{eqnarray}
 0  =   \partial_t \rho_1^+
 = (  q_1^{-} -  q_1^{+}    ) - (q_{3/2}^{+}  - q_{3/2}^{-} )  
 =   j_1  - j_{3/2}
\label{statioleft}
\end{eqnarray}
This is the empirical analog of Eq. \ref{steadyleft} concerning the steady state.

(iii) The summation of Eq. \ref{contrainteqs} over the $(N-1)$ spins $(S_1,...,S_{N-2},S_{N-1})$
yields the stationarity constraint for the 1-spin density $\rho_{N}^{+}=1-\rho_N^-$ 
\begin{eqnarray}
 0  =    \partial_t \rho_N^+
= ( q_{N-1/2}^{+}   - q_{N-1/2}^{-} ) - (  q_N^{+} - q_N^{-} )
 =   j_{N-1/2}-  j_N
\label{statioright}
\end{eqnarray}
This is the empirical analog of Eq. \ref{steadyright} concerning the steady state.

The physical meaning of Eq \ref{statiobulk}
 is simply that the local empirical currents $j_{i+1/2}$ flowing through the bulk links for $i=1,..,N-1$
 take the same value $j$, and this value $j$ also 
 corresponds to the incoming current $j_1=(q_1^{-}  -  q_1^{+}    )$ produced by the left reservoir (Eq. \ref{statioleft})
 and to the outgoing current $j_N=(q_N^{+} - q_N^{-}) $ produced by the right reservoir (Eq. \ref{statioright}).
 In summary, the stationarity of the 1-spin empirical density $\rho_i^+$ for the $N$ sites $i=1,..,N$
 is ensured by the following constraint where the $(N+1)$ local empirical currents $j_.$ have to take the same value $j$
 \begin{eqnarray}
j =j_1= j_{3/2} = j_{5/2}  = ... = j_{N-1/2} = j_N
 \label{jlinkuniform}
\end{eqnarray}

\subsubsection{ Stationarity of the 2-spin empirical density $\rho_{..}^{..}$ : closing the hierarchy problem via the Mean-field approximation for $ \rho_{..}^{..}$}

Now one needs to ensure the stationarity of the 2-spin density $\rho_{i,i+1}^{S_i,S_{i+1}} $ of Eq. \ref{rho2def}.
However the summation of the stationarity constraint of Eq. \ref{contrainteqs} 
over the remaining $(N-2)$ spins $n \ne (i,i+1)$ involves local empirical observables of three consecutive spins.
More generally, the local projections of Eq. \ref{contrainteqs} 
do not give closed constraints for local empirical observables, but produce a whole hierarchy,
which is the analog of the hierarchy for the steady state as recalled after Eq. \ref{steadyright}.
For many-body dynamics satisfying detailed-balance that converge towards the equilibrium without any steady current, one can use the vanishing of the optimal values
of the empirical currents to obtain closed large deviations properties for the local empirical densities and activities,
as discussed in detail in \cite{c_east} with the application to the Kinetically-constrained East model.
For the present non-equilibrium exclusion models that do not satisfy detailed-balance and that have non-vanishing currents in the steady state, the simplest approximation to close the hierarchy problem
is the Mean-field approximation for the 2-spin empirical density $\rho_{..}^{..}$
\begin{eqnarray}
\left[ \rho_{i,i+1}^{S_i,S_{i+1}} \right]^{MF}  = \rho_i^{S_i} \rho_{i+1}^{S_i}
\label{rho2MF}
\end{eqnarray}
which is the direct analog of the Mean-field approximation of Eq. \ref{P2steadyMF}
for the 2-spins probabilities $P_{..}^{..} $ in the steady state.


\subsection{ Formulation of the Mean-field approximation for the large deviations of local empirical observables }

Putting everything together, 
one obtains that the Mean-field approximation of Eq. \ref{rho2MF} for the 2-spin empirical density $\rho_{..}^{..}$
produces the following large deviations 
for the probability to see the local empirical density $\rho^+_i=1-\rho_i^-$ (Eq. \ref{rho1norma1}), 
the local empirical activities $a_.$ of Eqs \ref{aj} and \ref{ajbords},
and the global empirical current $j$ flowing through the whole sample (Eq. \ref{jlinkuniform})
\begin{eqnarray}
P_T^{[2.5]MF} \left( \rho_.^+ ; j ;  ;a_.  \right)  \opsimeq_{T \to +\infty}  
  \ e^{\displaystyle  - T  I_{2.5}^{MF} \left( \rho_.^.  ; j ; a_.  \right)     }
\label{MF2.5}
\end{eqnarray}
where the rate function $I_{2.5}^{MF} \left( \rho_.^.  ; j ; a_.  \right)$ 
is obtained from the rate function $I_{2.5} \left( \rho_.^. ; \rho_{..}^{..} ;q_.^.  \right)  $
of Eq. \ref{rateempi}
via the parametrization of Eqs \ref{qaj} and \ref{qajLR} for the flows $q_.^.$
and the Mean-field approximation of Eq. \ref{rho2MF} for the 2-spin empirical density $\rho_{..}^{..}$
\begin{eqnarray}
 I^{MF}_{2.5}[ \rho_.^+ ; j ; a_. ]
&& =   \sum_{i=1}^{N-1} 
\bigg[  \frac{ a_{i+1/2} + j}{2}    \ln \left( \frac{a_{i+1/2} + j }{  2 w_{i+1/2}^{+}   \rho_i^+ (1-\rho_{i+1}^+) }  \right) 
+ \frac{ a_{i+1/2}-j}{2}    \ln \left( \frac{a_{i+1/2}-j }{  2 w_{i+1/2}^{-}  (1- \rho_i^+) \rho_{i+1}^+ }  \right) 
\nonumber \\
&& \ \ \ \ \ \ \ \ \ \ \ -   a_{i+1/2}    + w_{i+1/2}^{+}  \rho_i^+ (1-\rho_{i+1}^+)  + w_{i+1/2}^{-}  (1-\rho_i^+) \rho_{i+1}^+  \bigg]
\nonumber \\
&& +  
  \frac{ a_1-j}{2}    \ln \left( \frac{a_1-j   }{ 2 w_1^{+}  \rho_1^+ } \right) 
 +   \frac{ a_1+ j}{2}   \ln \left( \frac{a_1+ j   }{ 2 w_1^{-}  (1-\rho_1^+)  } \right) 
 -  a_1     + w_1^{+}  \rho_1^+    + w_1^{-}  (1-\rho_1^+)   
 \nonumber \\
&& +  
\frac{ a_N+j}{2}  \ln \left( \frac{a_N+j  }{ 2 w_N^{+}  \rho_N^+ } \right) 
+ \frac{ a_N -  j}{2}   \ln \left( \frac{a_N -  j  }{ 2 w_N^{-}   (1-\rho_N^+) } \right) 
 - a_N  + w_N^{+}   \rho_N^+     + w_N^{-}   (1-\rho_N^+)  
\label{rate2.5MFaj}
\end{eqnarray}

Eq. \ref{MF2.5} characterizes how rare it is for large $T$  to see 
empirical observables  $[ \rho_.^+ ; j ;  a_.  ]  $
that are different from their typical values
given by the steady state values $ [ P_.^+ ; J ;  A_.  ] $ in the Mean-Field approximation of Eqs \ref{steadybulk}
\ref{steadyleft}  \ref{steadyright} and \ref{typlocalq} :
the Mean-Field steady state activities are given by
\begin{eqnarray}
A_{i+1/2}  && =  w_{i+1/2}^{+}   P_i^+(1-P_{i+1}^+)  +   w_{i+1/2}^{-}  (1-P_i^+) P_{i+1}^{+}
\nonumber \\
A_1  && =  w_1^{+}  P_1^+ +  w_1^{-} (1-P_1^+)
\nonumber \\
A_N  && =   w_N^{+}  P_{N}^+ +  w_N^{-}   (1-P_{N}^+)
\label{steadyactivities}
\end{eqnarray}
while the Mean-Field steady state current $J$ flowing through the whole sample reads
 \begin{eqnarray}
J   = w_{i+1/2}^{+}  P_i^+(1-P_{i+1}^+)  - w_{i+1/2}^{-} (1-P_i^+) P_{i+1}^{+}
= w_1^{-} (1-P_1^+) - w_1^{+}  P_1^+ = w_N^{+}  P_{N}^+   -  w_N^{-}   (1-P_{N}^+)
 \label{jsteady}
\end{eqnarray}


\subsection{ Time-additive observables involving only the 1-spin empirical density $\rho_.^.$
and the local empirical flows $q_.^.$ }

The large deviations of Eq. \ref{MF2.5} are interesting on their own as discussed above,
but they are also useful to analyze all the time-additive observables $O_T$ 
that can be written in terms of
 the 1-spin empirical density $ \rho_i^{\pm} $ and the local empirical flows $q_.^{\pm}$ 
 using some coefficients $(\alpha_.^{\pm} ; \beta_.^{\pm})$
\begin{eqnarray}
 O_T 
 =   \sum_{i=1}^N \left( \alpha_i^+  \rho_i^+ +\alpha_i^-  \rho_i^- \right)
 +   \sum_{i=1}^{N-1} \left( \beta_{i+1/2}^+  q_{i+1/2}^+ + \beta_{i+1/2}^-  q_{i-1/2}^+\right)
+ \beta_1^+ q_1^+ + \beta_1^- q_1^- 
+ \beta_N^+  q_N^+ + \beta_N^-  q_N^-
\label{additiveex}
\end{eqnarray}
Using the normalization of Eq. \ref{rho1norma1} to eliminate $\rho_i^-=1-\rho_i^+$
and the parametrization of Eqs \ref{qaj} and \ref{qajLR} for the empirical flows,
this observable can be rewritten in terms of the local densities $\rho_.^+$, of the local activities $a_.$
and of the global empirical current $j$ 
\begin{eqnarray}
 O_T 
  =  \alpha_0 +  \sum_{i=1}^N  \alpha_i  \rho_i^{+} 
  + \sum_{i=1}^{N-1}  \beta_{i+1/2} a_{i+1/2}
  +   \beta_1 a_1
 + \beta_N a_N
 +\nu j 
\label{additiveexaj}
\end{eqnarray}
with the appropriate coefficients
\begin{eqnarray}
\alpha_0   && =    \sum_{i=1}^N \alpha_i^{-} 
 \nonumber \\
\alpha_{i} && \equiv \alpha_i^{+} - \alpha_i^{-} 
 \nonumber \\
 \beta_{i+1/2} && \equiv \frac{   \beta_{i+1/2}^{+}+  \beta_{i+1/2}^{-}}{2} 
 \nonumber \\
  \beta_1 && \equiv \frac{   \beta_1^{+}+  \beta_1^{-}}{2} 
\nonumber \\
\beta_N && \equiv \frac{   \beta_N^{+}+  \beta_N^{-}}{2} 
\nonumber \\
\nu && \equiv
 \sum_{i=1}^{N-1}    \frac{   \beta_{i+1/2}^{+}-  \beta_{i+1/2}^{-}}{2}  +  \frac{   \beta_1^{-}-  \beta_1^{+}}{2} 
 + \frac{   \beta_N^{+}-  \beta_N^{-}}{2} 
\label{additiveexajcoefs}
\end{eqnarray}


\section{ Large deviations for inhomogeneous ASEP in the Mean-Field approximation }

\label{sec_asep}


In this section, we discuss the properties and the consequences
of the Large deviations at Level 2.5 in the Mean-Field approximation
of Eqs \ref{MF2.5} and \ref{rate2.5MFaj} for a given inhomogeneous ASEP sample defined by the rates $w_.^.$.

\subsection{ Gallavotti-Cohen symmetry of the Mean-Field rate function $ I^{MF}_{2.5}[ \rho_.^+ ; j ; a_. ] $ 
for opposite values $(j,-j)$  }

When the empirical density $\rho_.^{+}$ and the empirical activities $ a_. $ are given,
the difference of the rate function of Eq. \ref{rate2.5MFaj} for two opposite values $(\pm j)$ of the empirical current 
 \begin{eqnarray}
 I^{MF}_{2.5}[ \rho_.^+ ; j ; a_. ]- I^{MF}_{2.5}[ \rho_.^+ ; - j ; a_. ]
&& =  j \left[  \sum_{i=1}^{N-1} \ln \left( \frac{w_{i+1/2}^{-}  (1- \rho_i^+) \rho_{i+1}^+  }{ w_{i+1/2}^{+}   \rho_i^+ (1-\rho_{i+1}^+)  } \right)
 +  \ln \left( \frac{w_1^{+}  \rho_1^+  }{  w_1^{-}  (1-\rho_1^+)  } \right) 
 +    \ln \left( \frac{w_N^{-}   (1-\rho_N^+)  }{  w_N^{+}  \rho_N^+ } \right) \right]
 \nonumber \\
&& =  j \ln \left[ \frac{ w_1^+ } {w_1^-  } \left( \prod_{i=1}^{N-1} \frac{w_{i+1/2}^{-}    }{ w_{i+1/2}^{+}    } \right)
\frac{  w_N^-} { w_N^+ }
 \right]
\label{rate2.5MFgalla}
\end{eqnarray}
is linear in $j$ and the factor $\ln \left[ \frac{ w_1^+ } {w_1^-  } \left( \prod_{i=1}^{N-1} \frac{w_{i+1/2}^{-}    }{ w_{i+1/2}^{+}    } \right) \frac{  w_N^-} { w_N^+ }  \right]$ measures the irreversibility of the dynamics  : 
this is an example of the Gallavotti-Cohen fluctuation relations
(see \cite{galla,kurchan_langevin,Leb_spo,maes1999,jepps,derrida-lecture,harris_Schu,kurchan,searles,zia,chetrite_thesis,maes2009,maes2017,chetrite_HDR} and references therein).

Note that for the Symmetric model (SEP), 
where the two rates on each bulk link coincide $w_{i+1/2}^{\pm} = D_{i+1/2} $ (Eq. \ref{sep}),
the Gallavotti-Cohen symmetry of Eq. \ref{rate2.5MFgalla}
reduces to
 \begin{eqnarray}
(SEP) \ \ \ \ \  I^{MF}_{2.5}[ \rho_.^+ ; j ; a_. ]- I^{MF}_{2.5}[ \rho_.^+ ; - j ; a_. ]
 =  j \ln \left[ \frac{ w_1^+ w_N^-} {w_1^-  w_N^+ } \right]
\label{rate2.5MFgallasep}
\end{eqnarray}
since the irreversibility of the dynamics comes only from the two boundary spins connected to the two reservoirs.


\subsection{ Explicit contraction of the Level 2.5 over the activities $a_.$ 
to obtain the Level 2.25 for $[ \rho_.^+ ; j ] $  }

As already stressed for many other Markov jump processes \cite{maes_canonical,c_ring,c_interactions,c_detailed,c_runandtumble},
the contraction over the activities can be implemented explicitly.
The optimization of the rate function of Eq. \ref{rate2.5MFaj} over the local empirical activities $a_. $
\begin{eqnarray}
0 && = \frac{ \partial I^{MF}_{2.5}[ \rho_.^+ ; j ; a_. ]  }{\partial a_{i+1/2}}  =   
 \frac{  1}{2}    \ln \left(  \frac{ a_{i+1/2}^2-  j^2  }{4 w_{i+1/2}^{+}  w_{i+1/2}^{-}  \rho_i^+ (1- \rho_i^+)\rho_{i+1}^+(1-\rho_{i+1}^+)   }  \right)
 \nonumber \\
 0 && = \frac{ \partial I^{MF}_{2.5}[ \rho_.^+ ; j ; a_. ]  }{\partial a_1} 
 =    \frac{  1}{2}    \ln \left(  \frac{ a_1^2-  j^2  }{4 w_1^{+}  w_1^{-}  \rho_1^+ (1- \rho_1^+)   }  \right)
  \nonumber \\
 0 && = \frac{ \partial I^{MF}_{2.5}[ \rho_.^+ ; j ; a_. ]  }{\partial a_N} 
 =    \frac{  1}{2}    \ln \left(  \frac{ a_N^2-  j^2  }{4 w_N^{+}  w_N^{-}  \rho_N^+ (1- \rho_N^+)   }  \right)
\label{rate2.5ajderia}
\end{eqnarray}
yields 
the optimal values 
\begin{eqnarray}
  a_{i+1/2}^{opt}&& = \sqrt{ j^2+ 4 w_{i+1/2}^{+}  w_{i+1/2}^{-}  \rho_i^+ (1- \rho_i^+)\rho_{i+1}^+(1-\rho_{i+1}^+)  }  
  \nonumber \\
  a_1^{opt} && = \sqrt{ j^2+ 4 w_1^{+}  w_1^{-}  \rho_1^+ (1- \rho_1^+)  }
   \nonumber \\
  a_N^{opt} && = \sqrt{ j^2+ 4 w_N^{+}  w_N^{-}  \rho_N^+ (1- \rho_N^+)}
\label{aioptfa}
\end{eqnarray}
that can be plugged into Eq. \ref{rate2.5MFaj}
to obtain the rate function at Level 2.25 
 \begin{eqnarray}
 I^{MF}_{2.25}[ \rho_.^+ ; j ] && = I^{MF}_{2.5}[ \rho_.^+ ; j ; a_.^{opt} ]
 \nonumber \\
&& =   \sum_{i=1}^{N-1} 
\bigg[  j   \ln \left( \frac{ j + \sqrt{ j^2+ 4 w_{i+1/2}^{+}  w_{i+1/2}^{-}  \rho_i^+ (1- \rho_i^+)\rho_{i+1}^+(1-\rho_{i+1}^+)  }   }{  2 w_{i+1/2}^{+}   \rho_i^+ (1-\rho_{i+1}^+) }  \right) 
\nonumber \\
&& \ \ \ \ \ \ \ \ \ \ \ -  \sqrt{ j^2+ 4 w_{i+1/2}^{+}  w_{i+1/2}^{-}  \rho_i^+ (1- \rho_i^+)\rho_{i+1}^+(1-\rho_{i+1}^+)  }  
   + w_{i+1/2}^{+}  \rho_i^+ (1-\rho_{i+1}^+)  + w_{i+1/2}^{-}  (1-\rho_i^+) \rho_{i+1}^+  \bigg]
\nonumber \\
&& +  
j  \ln \left( \frac{ j +  \sqrt{ j^2+ 4 w_1^{+}  w_1^{-}  \rho_1^+ (1- \rho_1^+)  }  }{ 2 w_1^{-}  (1-\rho_1^+)  } \right) 
 -   \sqrt{ j^2+ 4 w_1^{+}  w_1^{-}  \rho_1^+ (1- \rho_1^+)  }     + w_1^{+}  \rho_1^+    + w_1^{-}  (1-\rho_1^+)   
 \nonumber \\
&& +  
j  \ln \left( \frac{j +  \sqrt{ j^2+ 4 w_N^{+}  w_N^{-}  \rho_N^+ (1- \rho_N^+)} }{ 2 w_N^{+}  \rho_N^+ } \right) 
 -  \sqrt{ j^2+ 4 w_N^{+}  w_N^{-}  \rho_N^+ (1- \rho_N^+)}  + w_N^{+}   \rho_N^+     + w_N^{-}   (1-\rho_N^+)  
\label{rate2.25MF}
\end{eqnarray}
that governs the joint probability of 
the global empirical current $j$ and of
  the local empirical density $\rho_i^{+}$ 
\begin{eqnarray}
P^{MF[2.25]}_T [ \rho_.^+ ; j   ] \opsimeq_{T \to +\infty} 
e^{ \displaystyle - T I^{MF}_{2.25} [  \rho_.^+ ; j ]  }
\label{level2.25j}
\end{eqnarray}

The rate function at Level 2.25 inherits the Gallavotti-Cohen symmetry of Eq. \ref{rate2.5MFgalla}
for the rate function at Level 2.5 as follows :
when the empirical density $\rho_.^{+}$ is given,
the difference of the rate function of Eq. \ref{rate2.25MF} for two opposite values $(\pm j)$ of the empirical current 
displays the same linear behavior in $j$
 \begin{eqnarray}
 I^{MF}_{2.25}[ \rho_.^+ ; j  ]- I^{MF}_{2.5}[ \rho_.^+ ; - j  ]
 =  j \ln \left[ \frac{ w_1^+ } {w_1^-  } \left( \prod_{i=1}^{N-1} \frac{w_{i+1/2}^{-}    }{ w_{i+1/2}^{+}    } \right)
\frac{  w_N^-} { w_N^+ }
 \right]
\label{rate2.25MFgalla}
\end{eqnarray}

For zero empirical current $j=0$, the rate function of Eq. \ref{rate2.25MF} simplifies into
 \begin{eqnarray}
 I^{MF}_{2.25}[ \rho_.^+ ; j=0 ]  
&& =   \sum_{i=1}^{N-1} 
\bigg[    \sqrt{ w_{i+1/2}^{+}  \rho_i^+ (1-\rho_{i+1}^+)  } - \sqrt{ w_{i+1/2}^{-}  (1-\rho_i^+) \rho_{i+1}^+ } \bigg]^2
\nonumber \\
&&  + \left[ \sqrt{w_1^{+}  \rho_1^+  }  - \sqrt{ w_1^{-}  (1-\rho_1^+)   } \right]^2
 +  
 \left[ \sqrt{ w_N^{+}   \rho_N^+    }  - \sqrt{ w_N^{-}   (1-\rho_N^+)  } \right]^2
\label{rate2.25MFjzero}
\end{eqnarray}
while the tails for large currents $j \to \pm  \infty$ are given by the leading term
\begin{eqnarray}
 I^{MF}_{2.25}[ \rho_.^+ ; j=0 ]  
&& \opsimeq_{j \to \pm \infty}  (N+1) \vert j \vert \ln \vert j \vert  
\label{rate2.25jinfinity}
\end{eqnarray}


\subsection{ Implicit contraction of the Level 2.25 over the current $j$ to obtain the Level 2 for the density $\rho_.^+$ }

The optimization of the rate function at Level 2.25 of Eq. \ref{rate2.25MF} over the global empirical current $j$
\begin{eqnarray}
0  = \frac{ \partial  I^{MF}_{2.25}[ \rho_.^+ ; j ]    } { \partial j }
&&  =   
 \sum_{i=1}^{N-1} 
     \ln \left( \frac{ j + \sqrt{ j^2+ 4 w_{i+1/2}^{+}  w_{i+1/2}^{-}  \rho_i^+ (1- \rho_i^+)\rho_{i+1}^+(1-\rho_{i+1}^+)  }   }
     {  2 w_{i+1/2}^{+}   \rho_i^+ (1-\rho_{i+1}^+) }  \right) 
\nonumber \\
&& +    \ln \left( \frac{ j +  \sqrt{ j^2+ 4 w_1^{+}  w_1^{-}  \rho_1^+ (1- \rho_1^+)  }  }{ 2 w_1^{-}  (1-\rho_1^+)  } \right) 
 +  \ln \left( \frac{j +  \sqrt{ j^2+ 4 w_N^{+}  w_N^{-}  \rho_N^+ (1- \rho_N^+)} }{ 2 w_N^{+}  \rho_N^+ } \right) 
  \label{rate2.25jderij}
\end{eqnarray}
yields that the optimal value $j_{opt}[\rho_.^+]$ as a function of the empirical density $\rho^+_i$ for $i=1,..,N$
is the solution of
\begin{eqnarray}
1   && = \left( \frac{ j_{opt}[\rho_.^+] +  \sqrt{ j_{opt}^2[\rho_.^+]+ 4 w_1^{+}  w_1^{-}  \rho_1^+ (1- \rho_1^+)  }  }{ 2 w_1^{-}  (1-\rho_1^+)  } \right) 
  \left( \frac{j_{opt}[\rho_.^+] +  \sqrt{ j_{opt}^2[\rho_.^+]+ 4 w_N^{+}  w_N^{-}  \rho_N^+ (1- \rho_N^+)} }{ 2 w_N^{+}  \rho_N^+ } \right) 
 \nonumber \\
 &&  \prod_{i=1}^{N-1} 
      \left( \frac{ j_{opt}[\rho_.^+] + \sqrt{ j_{opt}^2[\rho_.^+]+ 4 w_{i+1/2}^{+}  w_{i+1/2}^{-}  \rho_i^+ (1- \rho_i^+)\rho_{i+1}^+(1-\rho_{i+1}^+)  }   }
     {  2 w_{i+1/2}^{+}   \rho_i^+ (1-\rho_{i+1}^+) }  \right)
\label{jopt} 
\end{eqnarray}
One needs to plug this solution into Eq. \ref{rate2.25MF} to obtain the rate function at Level 2 for the empirical density alone
 \begin{eqnarray}
 I^{MF}_{2}[ \rho_.^+  ] && = I^{MF}_{2.25}[ \rho_.^+ ; j_{opt}[\rho_.^+] ] 
 \label{rate2MF} \\
&& =   \sum_{i=1}^{N-1} 
\bigg[  w_{i+1/2}^{+}  \rho_i^+ (1-\rho_{i+1}^+)  + w_{i+1/2}^{-}  (1-\rho_i^+) \rho_{i+1}^+
 -  \sqrt{ j_{opt}^2[\rho_.^+]+ 4 w_{i+1/2}^{+}  w_{i+1/2}^{-}  \rho_i^+ (1- \rho_i^+)\rho_{i+1}^+(1-\rho_{i+1}^+)  }  
     \bigg]
\nonumber \\
&& +   w_1^{+}  \rho_1^+    + w_1^{-}  (1-\rho_1^+)   
 -   \sqrt{ j_{opt}^2[\rho_.^+]+ 4 w_1^{+}  w_1^{-}  \rho_1^+ (1- \rho_1^+)  }     
 \nonumber \\ &&
 + w_N^{+}   \rho_N^+     + w_N^{-}   (1-\rho_N^+)    
 -  \sqrt{ j_{opt}^2[\rho_.^+]+ 4 w_N^{+}  w_N^{-}  \rho_N^+ (1- \rho_N^+)}  
\nonumber 
\end{eqnarray}
So here the Level 2 is not fully explicit as a consequence
 of Eq. \ref{jopt}  for the optimal current $j_{opt}[\rho_.^+] $,
 in contrast to the Totally Asymmetric model (TASEP) that will be discussed in subsection \ref{subsec_taseplevel2}.


\subsection{ Typical fluctuations of order $\frac{1}{\sqrt{T} }$ for the empirical densities and flows around their steady state values   }

\label{subsec_typfluct}

If one is interested only in the small typical fluctuations of order $\frac{1}{\sqrt{T} }$ around
the steady state values $ [ P_.^+ ; J ;  A_.  ] $ discussed in Eqs \ref{steadyactivities} and \ref{jsteady}
\begin{eqnarray}
 \rho_{i}^{+} && = P_{i}^{+} + \frac{{\hat \rho}_i^{+}}{\sqrt{T} }
\nonumber \\
 j && = J + \frac{{\hat j}}{\sqrt{T} }
\nonumber \\
 A_{i+1/2}  && =  A_{i+1/2}+  \frac{{\hat a}_{i+1/2}}{\sqrt{T} }
\nonumber \\
a_1 && =  A_1+ \frac{{\hat a}_1}{\sqrt{T} }
\nonumber \\
a_N && =  A_N + \frac{{\hat a}_N}{\sqrt{T} }
\label{hatrhoq}
\end{eqnarray}
one needs to expand
the rate function $I^{MF}_{2.5}[ \rho_.^+ ; j ; a_. ] $ of Eq. \ref{rate2.5MFaj}
at second order in the perturbations in order to obtain
the rescaled Gaussian rate function 
for the rescaled empirical observables $[{\hat \rho}_.^+ ; {\hat j} ; {\hat a}_. ]$
 \begin{eqnarray}
&& {\hat I}^{Gauss}_{2.5}[ {\hat \rho}_.^+ ; {\hat j} ; {\hat a}_. ]
 \equiv  \lim_{T \to + \infty}
 \left( T I_{2.5}^{MF} [  \rho_.^+=P_.^+ + \frac{{\hat \rho}_.^{+}}{\sqrt{T} } ; j = J + \frac{{\hat j}}{\sqrt{T} }; 
 a_.= A_.+  \frac{{\hat a}_.}{\sqrt{T} }  ] \right)
\nonumber \\ 
&& =    \sum_{i=1}^{N-1} 
\left(  \frac{  \left[ \frac{{\hat a}_{i+1/2}+ {\hat j}}{2} - w_{i+1/2}^{+} [ {\hat \rho}_i^+(1-P_{i+1}^+) - P_i {\hat \rho}_{i+1}^+ ]
\right]^2  }{ w_{i+1/2}^{+}   P_i^+(1-P_{i+1}^+)   } 
+\frac{  \left[ \frac{{\hat a}_{i+1/2}- {\hat j}}{2} - w_{i+1/2}^{-}
[ -{\hat \rho}_i^+ P_{i+1} + (1-P_i) {\hat \rho}_{i+1}^+]
\right]^2  }{   w_{i+1/2}^{-}  (1-P_i^+) P_{i+1}^{+}  } 
\right)
\nonumber \\
&& +  
\frac{  \left[ \frac{{\hat a}_1- {\hat j}}{2} - w_1^{+}{\hat \rho}_1^{+}\right]^2  }{  w_1^{+}  P_1^+  } 
+ \frac{  \left[ \frac{{\hat a}_1+ {\hat j}}{2} + w_1^{-}{\hat \rho}_1^{+}\right]^2  }{  w_1^{-} (1-P_1^+)  } 
 +  
\frac{  \left[ \frac{{\hat a}_N+ {\hat j}}{2} - w_N^{+}{\hat \rho}_N^{+}\right]^2  }{  w_N^{+}  P_{N}^+  } 
+  
\frac{  \left[ \frac{{\hat a}_N- {\hat j}}{2} + w_N^{-}{\hat \rho}_N^{+}\right]^2  }{  w_N^{-}  (1-P_{N}^+)  } 
\label{rate2.5MFajgauss}
\end{eqnarray}
that will govern the joint probability 
${\hat P}^{[2.5]}_T [{\hat \rho}_.^+ ; {\hat j} ; {\hat a}_. ] $ of the rescaled fluctuations $[{\hat \rho}_.^+ ; {\hat j} ; {\hat a}_. ] $
\begin{eqnarray}
{\hat P}^{[2.5]}_T [{\hat \rho}_.^+ ; {\hat j} ; {\hat a}_. ] \opsimeq_{T \to +\infty} e^{ - {\hat I}^{Gauss}_{2.5}[ {\hat \rho}_.^+ ; {\hat j} ; {\hat a}_. ]}
\label{level2.5gauss}
\end{eqnarray}

If one is only interested into the rescaled fluctuations $[{\hat \rho}_.^+ ; {\hat j} ]$ of the empirical density and of the current,
the corresponding rescaled Gaussian rate function reads
 \begin{eqnarray}
 {\hat I}^{Gauss}_{2.25}[ {\hat \rho}_.^+ ; {\hat j} ]
 && \equiv  \lim_{T \to + \infty}
  \left( T I_{2.25}^{MF} [  \rho_.^+=P_.^+ + \frac{{\hat \rho}_.^{+}}{\sqrt{T} } ; j = J + \frac{{\hat j}}{\sqrt{T} }  ] \right)
 \nonumber \\ 
&& =    \sum_{i=1}^{N-1} 
  \frac{  \left[ {\hat j} - 
  \left( {\hat \rho}_i^+ [w_{i+1/2}^{+} (1-P_{i+1}^+) - w_{i+1/2}^{-} P_{i+1} ] 
    + {\hat \rho}_{i+1}^+ [ - w_{i+1/2}^{+} P_i + w_{i+1/2}^{-}  (1-P_i) ] \right)
\right]^2  }
{ w_{i+1/2}^{+}   P_i^+(1-P_{i+1}^+) + w_{i+1/2}^{-}  (1-P_i^+) P_{i+1}^{+}   } 
\nonumber \\
&& 
+ \frac{  \left[  {\hat j} + \left( w_1^{+} + w_1^{-}\right) {\hat \rho}_1^{+}\right]^2  }
{ w_1^{-} + ( w_1^{+}  - w_1^{-} )P_1^+  } 
 +  \frac{  \left[  {\hat j} - \left( w_N^{+} +w_N^{-} \right){\hat \rho}_N^{+} \right]^2  }
{ w_N^{-} + ( w_N^{+}  - w_N^{-} )P_{N}^+ } 
\label{rate2.25MFjgauss}
\end{eqnarray}

The optimization of this rate function over the rescaled fluctuation $ {\hat j} $ of the current
 \begin{eqnarray}
0 && = \frac{1}{2}\frac{\partial  {\hat I}^{Gauss}_{2.25}[ {\hat \rho}_.^+ ; {\hat j} ] }{\partial  {\hat j}}
\nonumber \\
&& =  {\hat j} 
\left[  \sum_{i=1}^{N-1} 
  \frac{  1  } { w_{i+1/2}^{+}   P_i^+(1-P_{i+1}^+) + w_{i+1/2}^{-}  (1-P_i^+) P_{i+1}^{+}   } 
+  \frac{ 1  }{ w_1^{-} + ( w_1^{+}  - w_1^{-} )P_1^+  } 
 +  \frac{ 1  }{ w_N^{-} + ( w_N^{+}  - w_N^{-} )P_{N}^+ } \right]
\nonumber \\
&& -
  \sum_{i=1}^{N-1} 
  \frac{ 
  \left( {\hat \rho}_i^+ [w_{i+1/2}^{+} (1-P_{i+1}^+) - w_{i+1/2}^{-} P_{i+1} ] 
    + {\hat \rho}_{i+1}^+ [ - w_{i+1/2}^{+} P_i + w_{i+1/2}^{-}  (1-P_i) ] \right)  }
{ w_{i+1/2}^{+}   P_i^+(1-P_{i+1}^+) + w_{i+1/2}^{-}  (1-P_i^+) P_{i+1}^{+}   } 
\nonumber \\
&& 
+  \frac{   \left( w_1^{+} + w_1^{-}\right) {\hat \rho}_1^{+}  }
{ w_1^{-} + ( w_1^{+}  - w_1^{-} )P_1^+  } 
 -  \frac{   \left( w_N^{+} +w_N^{-} \right){\hat \rho}_N^{+}  }
{ w_N^{-} + ( w_N^{+}  - w_N^{-} )P_{N}^+ } 
\label{rate2.25MFjgaussderij}
\end{eqnarray}
yields the explicit optimal value (in contrast to the implicit Eq. \ref{jopt})
\begin{scriptsize}
 \begin{eqnarray}
  {\hat j}^{opt} [{\hat \rho}_.^+]
 = \frac{
\displaystyle  \sum_{i=1}^{N-1} 
  \frac{ 
  \left( {\hat \rho}_i^+ [w_{i+1/2}^{+} (1-P_{i+1}^+) - w_{i+1/2}^{-} P_{i+1} ] 
    + {\hat \rho}_{i+1}^+ [ - w_{i+1/2}^{+} P_i + w_{i+1/2}^{-}  (1-P_i) ] \right)  }
{ w_{i+1/2}^{+}   P_i^+(1-P_{i+1}^+) + w_{i+1/2}^{-}  (1-P_i^+) P_{i+1}^{+}   }  
-  \frac{   \left( w_1^{+} + w_1^{-}\right) {\hat \rho}_1^{+}  }
{ w_1^{-} + ( w_1^{+}  - w_1^{-} )P_1^+  } 
 +  \frac{   \left( w_N^{+} +w_N^{-} \right){\hat \rho}_N^{+}  }
{ w_N^{-} + ( w_N^{+}  - w_N^{-} )P_{N}^+ } 
}
{ \displaystyle
\sum_{i=1}^{N-1} 
  \frac{  1  } { w_{i+1/2}^{+}   P_i^+(1-P_{i+1}^+) + w_{i+1/2}^{-}  (1-P_i^+) P_{i+1}^{+}   } 
+  \frac{ 1  }{ w_1^{-} + ( w_1^{+}  - w_1^{-} )P_1^+  } 
 +  \frac{ 1  }{ w_N^{-} + ( w_N^{+}  - w_N^{-} )P_{N}^+ } 
}
\label{jopthat}
\end{eqnarray}
\end{scriptsize}
that can be plugged into Eq. \ref{rate2.25MFjgauss}
to obtain the rescaled Gaussian rate function for
the rescaled fluctuations of the empirical density ${\hat \rho}_.^+  $alone
 \begin{eqnarray}
{\hat I}^{Gauss}_{2}[ {\hat \rho}_.^+ ] && =  {\hat I}^{Gauss}_{2.25}[ {\hat \rho}_.^+ ; {\hat j}^{opt}  [{\hat \rho}_.^+]]
 \nonumber \\ 
&& =    \sum_{i=1}^{N-1} 
  \frac{  \left[ {\hat j}^{opt} [{\hat \rho}_.^+] - 
  \left( {\hat \rho}_i^+ [w_{i+1/2}^{+} (1-P_{i+1}^+) - w_{i+1/2}^{-} P_{i+1} ] 
    + {\hat \rho}_{i+1}^+ [ - w_{i+1/2}^{+} P_i + w_{i+1/2}^{-}  (1-P_i) ] \right)
\right]^2  }
{ w_{i+1/2}^{+}   P_i^+(1-P_{i+1}^+) + w_{i+1/2}^{-}  (1-P_i^+) P_{i+1}^{+}   } 
\nonumber \\
&& 
+ \frac{  \left[  {\hat j}^{opt} [{\hat \rho}_.^+] + \left( w_1^{+} + w_1^{-}\right) {\hat \rho}_1^{+}\right]^2  }
{ w_1^{-} + ( w_1^{+}  - w_1^{-} )P_1^+  } 
 +  \frac{  \left[  {\hat j}^{opt} [{\hat \rho}_.^+] - \left( w_N^{+} +w_N^{-} \right){\hat \rho}_N^{+} \right]^2  }
{ w_N^{-} + ( w_N^{+}  - w_N^{-} )P_{N}^+ } 
\label{rate2.2MFgauss}
\end{eqnarray}


\subsection{ Application to the large deviations of time-additive observables involving the local empirical observables  }

The time-additive observable $O_T$ of Eq. \ref{additiveex} involving only the 1-spin empirical density $\rho_.^{\pm}$ and the local empirical flows $q_.^{\pm}$ has been rewritten in terms of the local densities $\rho_.^+$, the local activities $a_.$
and the global empirical current $j$ in Eq. \ref{additiveexaj}.
As a consequence, its generating function of Eq. \ref{geneO} can be evaluated
from the Level 2.5 of Eq. \ref{MF2.5} via an integral over the empirical variables $ [ \rho_.^+ ; j ;  a_.   ] $
\begin{eqnarray}
\langle e^{ T k O } \rangle 
 && = \int_{-\infty}^{+\infty} dj \left[ \prod_{i=1}^N \int_0^1 d \rho_i^+ \right]  \left[ \prod_{i=1}^{N-1} 
 \int_0^{+\infty} d a_{i+1/2} \right] \int_0^{+\infty} d a_1 \int_0^{+\infty} d a_N 
\nonumber \\
&&  P^{MF[2.5]}_T [ \rho_.^+ ; j ;  a_.   ]
 e^{ \displaystyle T k  \left[ \alpha_0 +  \sum_{i=1}^N  \alpha_i  \rho_i^{+} 
  + \sum_{i=1}^{N-1}  \beta_{i+1/2} a_{i+1/2}  +   \beta_1 a_1 + \beta_N a_N +\nu j \right] }
\nonumber \\
&& \opsimeq_{T \to +\infty} 
 \int_{-\infty}^{+\infty} dj \left[ \prod_{i=1}^N \int_0^1 d \rho_i^+ \right]  \left[ \prod_{i=1}^{N-1} 
 \int_0^{+\infty} d a_{i+1/2} \right] \int_0^{+\infty} d a_1 \int_0^{+\infty} d a_N 
e^{ \displaystyle  - T L_{2.5}^{[k]}  [  \rho_.^+ ; j ; a_. ]  }
\label{geneO2.5}
\end{eqnarray}
with the function
\begin{eqnarray}
 L_{2.5}^{[k]}  [  \rho_.^+ ; j ; a_. ]  && =   I^{MF}_{2.5} [  \rho_.^+ ; j ; a_. ] - k  \left[ \alpha_0 +  \sum_{i=1}^N  \alpha_i  \rho_i^{+} 
  + \sum_{i=1}^{N-1}  \beta_{i+1/2} a_{i+1/2}  +   \beta_1 a_1 + \beta_N a_N +\nu j \right]
\label{Lk2.5}
\end{eqnarray}
The optimization over the local empirical activities $a_. $
\begin{eqnarray}
0 && = \frac{ \partial L_{2.5}^{[k]}[ \rho_.^+ ; j ; a_. ]  }{\partial a_{i+1/2}}  =   
 \frac{  1}{2}    \ln \left(  \frac{ a_{i+1/2}^2-  j^2  }{4 w_{i+1/2}^{+}  w_{i+1/2}^{-}  \rho_i^+ (1- \rho_i^+)\rho_{i+1}^+(1-\rho_{i+1}^+)   }  \right) - k \beta_{i+1/2}
 \nonumber \\
 0 && = \frac{ \partial L_{2.5}^{[k]}[ \rho_.^+ ; j ; a_. ]  }{\partial a_1} 
 =    \frac{  1}{2}    \ln \left(  \frac{ a_1^2-  j^2  }{4 w_1^{+}  w_1^{-}  \rho_1^+ (1- \rho_1^+)   }  \right) - k  \beta_1
  \nonumber \\
 0 && = \frac{ \partial L_{2.5}^{[k]}[ \rho_.^+ ; j ; a_. ]  }{\partial a_N} 
 =    \frac{  1}{2}    \ln \left(  \frac{ a_N^2-  j^2  }{4 w_N^{+}  w_N^{-}  \rho_N^+ (1- \rho_N^+)   }  \right)- k  \beta_N
\label{zk2.5ajderia}
\end{eqnarray}
leads to the optimal values 
\begin{eqnarray}
  a_{i+1/2}^{opt}&& = \sqrt{ j^2+ 4 e^{2k \beta_{i+1/2}} w_{i+1/2}^{+}  w_{i+1/2}^{-}  \rho_i^+ (1- \rho_i^+)\rho_{i+1}^+(1-\rho_{i+1}^+)  }  
  \nonumber \\
  a_1^{opt} && = \sqrt{ j^2+ 4 e^{2k \beta_1}w_1^{+}  w_1^{-}  \rho_1^+ (1- \rho_1^+)  }
   \nonumber \\
  a_N^{opt} && = \sqrt{ j^2+ 4 e^{2k \beta_N}w_N^{+}  w_N^{-}  \rho_N^+ (1- \rho_N^+)}
\label{aioptfazk}
\end{eqnarray}
that can be plugged into Eq. \ref{Lk2.5} to obtain the function of the empirical density $\rho_.^+$ and of the current $j$
\begin{eqnarray}
&& L_{2.25}^{[k]}  [  \rho_.^+ ; j  ]   =  L_{2.5}^{[k]}  [  \rho_.^+ ; j ; a_.^{opt} ]  
 \nonumber \\
 && = 
 \sum_{i=1}^{N-1} 
\bigg[  j   \ln \left( \frac{ j + \sqrt{ j^2+ 4 e^{2k \beta_{i+1/2}}w_{i+1/2}^{+}  w_{i+1/2}^{-}  \rho_i^+ (1- \rho_i^+)\rho_{i+1}^+(1-\rho_{i+1}^+)  }   }{  2 e^{k \beta_{i+1/2}}w_{i+1/2}^{+}   \rho_i^+ (1-\rho_{i+1}^+) }  \right) 
\nonumber \\
&& \ \ \ \ \ \ \ \ \ \ \ -  \sqrt{ j^2+ 4 e^{2k \beta_{i+1/2}}w_{i+1/2}^{+}  w_{i+1/2}^{-}  \rho_i^+ (1- \rho_i^+)\rho_{i+1}^+(1-\rho_{i+1}^+)  }  
   + w_{i+1/2}^{+}  \rho_i^+ (1-\rho_{i+1}^+)  + w_{i+1/2}^{-}  (1-\rho_i^+) \rho_{i+1}^+  \bigg]
\nonumber \\
&& +  
j  \ln \left( \frac{ j +  \sqrt{ j^2+ 4 e^{2k \beta_1}w_1^{+}  w_1^{-}  \rho_1^+ (1- \rho_1^+)  }  }{ 2 e^{k \beta_1}w_1^{-}  (1-\rho_1^+)  } \right) 
 -   \sqrt{ j^2+ 4 e^{2k \beta_1}w_1^{+}  w_1^{-}  \rho_1^+ (1- \rho_1^+)  }     + w_1^{+}  \rho_1^+    + w_1^{-}  (1-\rho_1^+)   
 \nonumber \\
&& +  
j  \ln \left( \frac{j +  \sqrt{ j^2+ 4 e^{2k \beta_N}w_N^{+}  w_N^{-}  \rho_N^+ (1- \rho_N^+)} }{ 2 e^{k \beta_N}w_N^{+}  \rho_N^+ } \right) 
 -  \sqrt{ j^2+ 4 e^{2k \beta_N}w_N^{+}  w_N^{-}  \rho_N^+ (1- \rho_N^+)}  + w_N^{+}   \rho_N^+     + w_N^{-}   (1-\rho_N^+)   
 \nonumber \\
 &&+ k   \alpha_0 +  k \sum_{i=1}^N  \alpha_i  \rho_i^{+}  + k \nu j 
 \label{Lk2.25}
\end{eqnarray}
that governs the generating function of Eq. \ref{geneO2.5}
\begin{eqnarray}
\langle e^{ T k O } \rangle 
 &&  \opsimeq_{T \to +\infty} 
 \int_{-\infty}^{+\infty} dj \left[ \prod_{i=1}^N \int_0^1 d \rho_i^+ \right]   
e^{ \displaystyle  - T L_{2.25}^{[k]}  [  \rho_.^+ ; j  ]  } \opsimeq_{T \to +\infty} e^{ \displaystyle  T G(k)  }
\label{geneO2.25}
\end{eqnarray}
So the scaled cumulants generation function $G(k)$ of Eq. \ref{gkper}
corresponds to the optimization of $  \left( -  L_{2.25}^{[k]}  [  \rho_.^+ ; j  ] \right) $ 
over the $N$ values of the empirical density $\rho_.^+ $ for $i=1,..,N$ and over the empirical current $j$.

If one is interested only in the two first cumulants, one can use the analysis of the previous subsection \ref{subsec_typfluct} as follows.
As recalled in Appendix \ref{app_reminder2.5}, the first cumulant $G_1$ of Eq. \ref{g1} corresponds to the steady state value $O_{st}$
involving the steady state $P_.^+$, the steady activities $A_.$ and the steady current $J$
\begin{eqnarray}
G_1 = \langle O_T \rangle = \alpha_0 +  \sum_{i=1}^N  \alpha_i P_i^{+} 
  + \sum_{i=1}^{N-1}  \beta_{i+1/2} A_{i+1/2}  +   \beta_1 A_1 + \beta_N A_N +\nu J
\label{g1steady}
\end{eqnarray}
The small typical fluctuations of order $\frac{1}{\sqrt{T} } $ around this steady state value 
can be rewritten in terms of the rescaled empirical observables 
$[{\hat \rho}_.^+ ; {\hat j} ; {\hat a}_. ]$ of Eq. \ref{hatrhoq}
\begin{eqnarray}
O_T - \langle O_T \rangle
 = \frac{1}{\sqrt{T} }   \left( 
   \sum_{i=1}^N  \alpha_i  {\hat \rho}^{+}_i 
  + \sum_{i=1}^{N-1}  \beta_{i+1/2} {\hat a}_{i+1/2}  +   \beta_1 a_1 + \beta_N a_N +\nu {\hat j} \right)
\label{additivesmallfluct}
\end{eqnarray}
so that the rescaled variance of Eq. \ref{g2} 
\begin{eqnarray}
G_2  \equiv  T   \langle  \left( O_T  - \langle  O_T  \rangle \right)^2  \rangle
= \bigg\langle \left( \sum_{i=1}^N  \alpha_i  {\hat \rho}^{+}_i 
  + \sum_{i=1}^{N-1}  \beta_{i+1/2} {\hat a}_{i+1/2}  +   \beta_1 a_1 + \beta_N a_N +\nu {\hat j}  \right)^2  \bigg\rangle
\label{g2eval}
\end{eqnarray}
can be evaluated via the average over the Gaussian probability ${\hat P}^{[2.5]}_T [{\hat \rho}_.^+ ; {\hat j} ; {\hat a}_. ] $ of the rescaled fluctuations of Eq. \ref{level2.5gauss}.


\section{ Large deviations for inhomogeneous TASEP in the MF approximation }

\label{sec_tasep}

In this section, we describe the simplifications that occur for TASEP
with respect to the properties described in section \ref{sec_asep} for ASEP.

\subsection{ Rate function $ I^{MF}_{2.5}[ \rho_.^{+} ; j] $ in the Mean-field approximation }

For the inhomogeneous Totally Asymmetric model (TASEP) where the rates corresponding to backward motion vanish (Eq. \ref{tasep}),
the corresponding empirical flows also vanish
\begin{eqnarray}
 q_{i+1/2}^{-} && = 0 \ \ {\rm for } \ \ i=1,..,N-1
 \nonumber \\
 q_1^+ && =0
  \nonumber \\
 q_N^- && =0
\label{tasepq}
\end{eqnarray}
As a consequence, in the parametrization of Eq. \ref{qajLR},
the remaining non-vanishing flows coincide with the activities 
and they can all be rewritten in terms of the global empirical current $j$ of Eq. \ref{jlinkuniform} alone
\begin{eqnarray}
 q_{i +1/2}^+  && = a_{i +1/2} = j \ \ {\rm for } \ \ i=1,..,N-1
\nonumber \\
  q_1^- && = a_1 = j
    \nonumber \\
  q_N^+ && = a_N = j
 \label{tasepqj}
\end{eqnarray}
So the Mean-Field rate function at Level 2.5 of Eq. \ref{rate2.5MFaj} 
only involves the global empirical positive current $j \in [0,+\infty[$ and the 1-spin empirical density $\rho_i^+$ 
\begin{eqnarray}
 I^{MF}_{2.5}[ \rho_.^{+} ; j]
&& =   \sum_{i=1}^{N-1} 
\left[ j  \ln \left( \frac{j }
{  w_{i+1/2}^{+}   \rho_i^+ (1-\rho_{i+1}^+) }  \right) 
- j   + w_{i+1/2}^{+}  \rho_i^+ (1-\rho_{i+1}^+)  \right]
\nonumber \\
&&
 + 
 \left[ j \ln \left( \frac{j  }{  w_1^{-}  (1-\rho_1^+)  } \right) 
 - j  + w_1^{-}  (1-\rho_1^+ )  \right]
 +  
\left[ j \ln \left( \frac{j }{  w_N^{+}  \rho_N^+ } \right) 
 - j + w_N^{+}   \rho_N^+  \right]
\label{rate2.5MFtasep}
\end{eqnarray}
At zero empirical current $j=0$, this rate function reduces to
\begin{eqnarray}
 I^{MF}_{2.5}[ \rho_.^{+} ; j=0]
&& =   \sum_{i=1}^{N-1}  w_{i+1/2}^{+}  \rho_i^+ (1-\rho_{i+1}^+)  
  + w_1^{-}  (1-\rho_1^+ )  
 + w_N^{+}   \rho_N^+ 
\label{rate2.5directedzero}
\end{eqnarray}
while the tail for large current $j \to +\infty$ is governed by the leading term
\begin{eqnarray}
 I^{MF}_{2.5}[ \rho_.^{+} ; j] \opsimeq_{j \to +\infty} (N+1)  j \ln j 
\label{rate2.5directedinfinity}
\end{eqnarray}


\subsection{ Explicit contraction of the Level 2.5 over the global empirical current $j$ to obtain the Level 2 for $\rho_.^+ $ }

\label{subsec_taseplevel2}

The optimization of the rate function at Level 2.5 of Eq. \ref{rate2.5MFtasep} over the empirical global current $j$
\begin{eqnarray}
0  = \frac{ \partial  I^{MF}_{2.5}[ \rho_.^+ ; j ]    } { \partial j }
&&  =   
 \sum_{i=1}^{N-1}    \ln \left( \frac{j }{  w_{i+1/2}^{+}   \rho_i^+ (1-\rho_{i+1}^+) }  \right) 
+ \ln \left( \frac{j  }{  w_1^{-}  (1-\rho_1^+)  } \right) 
+  \ln \left( \frac{j }{  w_N^{+}  \rho_N^+ } \right) 
\nonumber \\
&& = (N+1) \ln(j) -  \sum_{i=1}^{N-1}    \ln \left(  w_{i+1/2}^{+}   \rho_i^+ (1-\rho_{i+1}^+)   \right) 
- \ln \left(  w_1^{-}  (1-\rho_1^+)  \right) 
-  \ln \left(   w_N^{+}  \rho_N^+  \right) 
  \label{rate2.5derijtasep}
\end{eqnarray}
yields the optimal value as a function of the empirical density $\rho_.^+$
\begin{eqnarray}
j_{opt} [\rho_.^+] =    \left( w_1^{-} w_N^{+} \left[ \prod_{n=1}^{N-1} w_{n+1/2}^{+} \right] \left[ \prod_{i=1}^{N} \rho_i^+ (1-\rho_i^+) \right] 
\right)^{  \frac{1}{N+1} } 
\label{jopttasep} 
\end{eqnarray}
that can be plugged into Eq. \ref{rate2.5MFtasep} to obtain the rate function at Level 2 for the empirical density $\rho_.^+$ alone
 \begin{eqnarray}
 && I^{MF}_{2}[ \rho_.^+  ]  = I^{MF}_{2.5}[ \rho_.^+ ; j_{opt}[\rho_.^+] ] 
 =   \sum_{i=1}^{N-1}  w_{i+1/2}^{+}  \rho_i^+ (1-\rho_{i+1}^+) 
 + w_1^{-}  (1-\rho_1^+ ) 
+ w_N^{+}   \rho_N^+
   - (N+1) j_{opt}[\rho_.^+]
   \nonumber \\
   && 
   =    \sum_{i=1}^{N-1}  w_{i+1/2}^{+}  \rho_i^+ (1-\rho_{i+1}^+) 
 + w_1^{-}  (1-\rho_1^+ ) 
+ w_N^{+}   \rho_N^+
   - (N+1)  \left( w_1^{-} w_N^{+} \left[ \prod_{n=1}^{N-1} w_{n+1/2}^{+} \right] \left[ \prod_{i=1}^{N} \rho_i^+ (1-\rho_i^+) \right] 
\right)^{  \frac{1}{N+1} } 
\label{rate2MFtasep}
\end{eqnarray}
As usually expected, the contraction over the global empirical current $j$ transforms 
the additive local functional $ I^{MF}_{2.5}[ \rho_.^{+} ; j] $ at Level 2.5
into a non-additive functional $  I^{MF}_{2}[ \rho_.^+  ] $ at Level 2.

To see more clearly the physical meaning,
one can use the steady state $P_.^+$ and the steady current $J$ of Eq. \ref{jsteady}
to rewrite the rates as
 \begin{eqnarray}
 w_{i+1/2}^{+}  && = \frac{ J}{P_i^+(1-P_{i+1}^+)  } \ \ \ {\rm for } \ \ i=1,..,N-1
 \nonumber \\
 w_1^{-} && = \frac{J}{(1-P_1^+) } 
  \nonumber \\
  w_N^{+} && = \frac{J}{ P_{N}^+   }
 \label{tasepjsteady}
\end{eqnarray}
Then the optimal current of Eq. \ref{jopttasep} 
becomes
\begin{eqnarray}
j_{opt} [\rho_.^+]  =  J   \left(  \prod_{i=1}^{N} \frac{ \rho_i^+ (1-\rho_i^+) }{ P_i^+ (1-P_i^+) }\right)^{  \frac{1}{N+1} } 
= J \ e^{ \displaystyle  \frac{1}{N+1} \sum_{i=1}^{N} \ln \left( \frac{ \rho_i^+ (1-\rho_i^+) }{ P_i^+ (1-P_i^+) } \right) } 
\label{jopttasepsteady} 
\end{eqnarray}
and the rate function of Eq. \ref{rate2MFtasep}
reads
 \begin{eqnarray}
  I^{MF}_{2}[ \rho_.^+  ]  
   =  J \left[   \sum_{i=1}^{N-1}   \frac{ \rho_i^+ (1-\rho_{i+1}^+) }{P_i^+(1-P_{i+1}^+)  }  
 + \frac{(1-\rho_1^+ ) }{(1-P_1^+) }   
+ \frac{\rho_N^+}{ P_{N}^+   }   
   - (N+1)\left(  \prod_{i=1}^{N} \frac{ \rho_i^+ (1-\rho_i^+) }{ P_i^+ (1-P_i^+) }\right)^{  \frac{1}{N+1} } 
   \right]
\label{rate2MFtasepsteady}
\end{eqnarray}


\subsection{ Typical fluctuations of order $\frac{1}{\sqrt{T} }$ for the empirical density and current around their steady state values   }

\label{subsec_typfluctTASEP}

If one is interested only in the small typical fluctuations of order $\frac{1}{\sqrt{T} }$ around the steady state values
$ [ P_.^+ ; J   ] $
\begin{eqnarray}
 \rho_{i}^{+} && = P_{i}^{+} + \frac{{\hat \rho}_i^{+}}{\sqrt{T} }
\nonumber \\
 j && = J + \frac{{\hat j}}{\sqrt{T} }
\label{hatrhoqtasep}
\end{eqnarray}
one needs to expand
the rate function $I^{MF}_{2.5}[ \rho_.^+ ; j ] $ of Eq. \ref{rate2.5MFtasep}
at second order in the perturbations to obtain
the rescaled Gaussian rate function for the rescaled empirical observables $[{\hat \rho}_.^+ ; {\hat j}  ]$
\begin{eqnarray}
 {\hat I}^{Gauss}_{2.5}[ {\hat \rho}_.^+ ; {\hat j} ]
  && \equiv  \lim_{T \to + \infty}
 \left( T I_{2.5}^{MF} [  \rho_.^+=P_.^+ + \frac{{\hat \rho}_.^{+}}{\sqrt{T} } ; j = J + \frac{{\hat j}}{\sqrt{T} }  ] \right)
\nonumber \\ 
&& =    \sum_{i=1}^{N-1} 
   \frac{\left(  {\hat j} - w_{i+1/2}^{+}      [ {\hat \rho}_i^+(1-P_{i+1}^+) - P_i^+ {\hat \rho}_{i+1}^+ ] \right)^2 }{  2 J } 
  +    \frac{\left(  {\hat j} + w_1^{-}  {\hat \rho} _1^+ \right)^2 }{  2 J } 
 +    \frac{\left(  {\hat j} - w_N^{+}  {\hat \rho} _N^+ \right)^2 }{  2 J } 
\label{rate2.5tasepgauss}
\end{eqnarray}
that will govern the joint probability 
${\hat P}^{[2.5]}_T [{\hat \rho}_.^+ ; {\hat j}  ] $ of the rescaled fluctuations $ [{\hat \rho}_.^+ ; {\hat j}  ]$
\begin{eqnarray}
{\hat P}^{[2.5]}_T [{\hat \rho}_.^+ ; {\hat j} ] \opsimeq_{T \to +\infty} 
e^{ - {\hat I}^{Gauss}_{2.5}[ {\hat \rho}_.^+ ; {\hat j} ]}
\label{level2.5gausstasep}
\end{eqnarray}

Using again Eq. \ref{tasepjsteady} to replace the rates in terms of the steady state $P_.^+$ and the steady current $J$,
Eq. \ref{rate2.5tasepgauss} can be rewritten as
\begin{eqnarray}
 {\hat I}^{Gauss}_{2.5}[ {\hat \rho}_.^+ ; {\hat j} ]
 =    \sum_{i=1}^{N-1} 
   \frac{\left(  {\hat j} - J    \left[ \frac{ {\hat \rho}_i^+ }{P_i^+  }   - \frac{ {\hat \rho}_{i+1}^+}{(1-P_{i+1}^+)  }  \right] \right)^2 }{  2 J } 
  +    \frac{\left(  {\hat j} + J \frac{{\hat \rho} _1^+}{(1-P_1^+) }    \right)^2 }{  2 J } 
 +    \frac{\left(  {\hat j} - J \frac{ {\hat \rho} _N^+ }{ P_{N}^+   }   \right)^2 }{  2 J } 
\label{rate2.5tasepgausssansw}
\end{eqnarray}

The optimization over the rescaled fluctuation $ {\hat j} $ of the current
 \begin{eqnarray}
0  =  \frac{\partial  {\hat I}^{Gauss}_{2.5}[ {\hat \rho}_.^+ ; {\hat j} ] }{\partial  {\hat j}}
 =  (N+1) \frac{ {\hat j} }{J} -  \sum_{i=1}^{N-1} 
 \left[ \frac{ {\hat \rho}_i^+ }{P_i^+  }   - \frac{ {\hat \rho}_{i+1}^+}{(1-P_{i+1}^+)  }  \right]
  +    \frac{{\hat \rho} _1^+}{(1-P_1^+) }  
  -  \frac{ {\hat \rho} _N^+ }{ P_{N}^+   } 
\label{rate2.5MFtase[gaussderij}
\end{eqnarray}
yields the optimal value
 \begin{eqnarray}
 {\hat j}^{opt} [ {\hat \rho}_.^+ ]  
 = \frac{J}{N+1}   \sum_{i=1}^{N} {\hat \rho}_i^+ \left(  \frac{ 1 }{P_i^+  }     -   \frac{ 1}{(1-P_{i}^+)  }  \right)  
 =  \frac{J}{N+1}   \sum_{i=1}^{N} {\hat \rho}_i^+ \left(  \frac{ 1- 2 P_i^+ }{P_i^+ (1-P_{i}^+) }   \right)  
\label{jopthattasep}
\end{eqnarray}
that can be plugged into Eq. \ref{rate2.5tasepgausssansw}
to obtain the rescaled Gaussian rate function for
the rescaled fluctuations of the empirical density ${\hat \rho}_.^+  $alone
 \begin{eqnarray}
{\hat I}^{Gauss}_{2}[ {\hat \rho}_.^+ ] && =  {\hat I}^{Gauss}_{2.5}[ {\hat \rho}_.^+ ; {\hat j}^{opt} [ {\hat \rho}_.^+ ]  ]
 \nonumber \\ 
&& =  
  \sum_{i=1}^{N-1} 
   \frac{\left(  {\hat j}^{opt} [ {\hat \rho}_.^+ ]  - J    \left[ \frac{ {\hat \rho}_i^+ }{P_i^+  }   - \frac{ {\hat \rho}_{i+1}^+}{(1-P_{i+1}^+)  }  \right] \right)^2 }{  2 J } 
  +    \frac{\left(  {\hat j}^{opt} [ {\hat \rho}_.^+ ]  + J \frac{{\hat \rho} _1^+}{(1-P_1^+) }    \right)^2 }{  2 J } 
 +    \frac{\left(  {\hat j}^{opt} [ {\hat \rho}_.^+ ]  - J \frac{ {\hat \rho} _N^+ }{ P_{N}^+   }   \right)^2 }{  2 J } 
\label{rate2.MFtasepgauss}
\end{eqnarray}



\subsection{ Application to the large deviations of time-additive observables involving the local empirical observables  }

For the Totally Asymmetric model, 
where the local empirical flows are given by Eqs \ref{tasepq} and \ref{tasepqj},
the time-additive observable $O_T$ of Eqs \ref{additiveex} and Eq. \ref{additiveexaj}
only involves the empirical density $\rho_.^+$ and the empirical current $j$
\begin{eqnarray}
 O_T && =    \alpha_0 +  \sum_{i=1}^N  \alpha_i  \rho_i^{+}  +\nu j 
\label{additivetasep}
\end{eqnarray}

As a consequence, its generating function of Eq. \ref{geneO} can be evaluated
from the Level 2.5 of Eq. \ref{rate2.5MFtasep} via an integral over the empirical variables $ [ \rho_.^+ ; j   ] $
\begin{eqnarray}
\langle e^{ T k O } \rangle 
 && = \left[ \prod_{i=1}^N \int_0^1 d \rho_i^+ \right] 
\int_{-\infty}^{+\infty} dj   P^{MF[2.5]}_T [ \rho_.^+ ; j   ]
 e^{ \displaystyle T k  \left[ \alpha_0 +  \sum_{i=1}^N  \alpha_i  \rho_i^{+}  +\nu j \right] }
\nonumber \\
&& \opsimeq_{T \to +\infty} 
  \left[ \prod_{i=1}^N \int_0^1 d \rho_i^+ \right]  
\int_{-\infty}^{+\infty} dj  e^{ \displaystyle  - T L_{2.5}^{[k]}  [  \rho_.^+ ; j  ]  }
\label{geneO2.5tasep}
\end{eqnarray}
with the function
\begin{eqnarray}
 L_{2.5}^{[k]}  [  \rho_.^+ ; j  ]  && =   I^{MF}_{2.5} [  \rho_.^+ ; j  ] 
 - k  \left[ \alpha_0 +  \sum_{i=1}^N  \alpha_i  \rho_i^{+}   +\nu j \right]
\label{Lk2.5tasep}
\end{eqnarray}
The optimization over the empirical global current $j$
\begin{eqnarray}
0  = \frac{ \partial L_{2.5}^{[k]}  [  \rho_.^+ ; j  ]    } { \partial j }
&&  =   
 \sum_{i=1}^{N-1}    \ln \left( \frac{j }{  w_{i+1/2}^{+}   \rho_i^+ (1-\rho_{i+1}^+) }  \right) 
+ \ln \left( \frac{j  }{  w_1^{-}  (1-\rho_1^+)  } \right) 
+  \ln \left( \frac{j }{  w_N^{+}  \rho_N^+ } \right) - k \nu 
\nonumber \\
&& = (N+1) \ln(j) -  \sum_{i=1}^{N-1}    \ln \left(  w_{i+1/2}^{+}   \rho_i^+ (1-\rho_{i+1}^+)   \right) 
- \ln \left(  w_1^{-}  (1-\rho_1^+)  \right) 
-  \ln \left(   w_N^{+}  \rho_N^+  \right) - k \nu 
  \label{l2.5derijtasep}
\end{eqnarray}
yields the optimal value as a function of the empirical density $\rho_.^+$ and of the parameter $k$
\begin{eqnarray}
j_{opt} [\rho_.^+,k] = e^{k \nu}   \left( w_1^{-} w_N^{+} \left[ \prod_{n=1}^{N-1} w_{n+1/2}^{+} \right] \left[ \prod_{i=1}^{N} \rho_i^+ (1-\rho_i^+) \right] 
\right)^{  \frac{1}{N+1} } 
\label{jopttasepk} 
\end{eqnarray}
that can be plugged into Eq. \ref{Lk2.5tasep} to obtain the function of the empirical density $\rho_.^+$ alone
 \begin{eqnarray}
 && L_{2}^{[k]}[ \rho_.^+  ]  = L_{2.5}^{[k]}  [  \rho_.^+ ; j_{opt}[\rho_.^+,k] ] 
\nonumber \\
&& =   \sum_{i=1}^{N-1}  w_{i+1/2}^{+}  \rho_i^+ (1-\rho_{i+1}^+) 
 + w_1^{-}  (1-\rho_1^+ ) 
+ w_N^{+}   \rho_N^+
   - (N+1) j_{opt}[\rho_.^+,k]
   - k  \left[ \alpha_0 +  \sum_{i=1}^N  \alpha_i  \rho_i^{+}    \right]
   \nonumber \\
   && 
   =    \sum_{i=1}^{N-1}  w_{i+1/2}^{+}  \rho_i^+ (1-\rho_{i+1}^+) 
 + w_1^{-}  (1-\rho_1^+ ) 
+ w_N^{+}   \rho_N^+
   - (N+1) e^{k \nu} \left( w_1^{-} w_N^{+} \left[ \prod_{n=1}^{N-1} w_{n+1/2}^{+} \right] \left[ \prod_{i=1}^{N} \rho_i^+ (1-\rho_i^+) \right] 
\right)^{  \frac{1}{N+1} } 
\nonumber \\
&& - k  \left[ \alpha_0 +  \sum_{i=1}^N  \alpha_i  \rho_i^{+}   \right]
\label{l2MFtasep}
\end{eqnarray}
that governs the generating function of Eq. \ref{geneO2.5tasep}
\begin{eqnarray}
\langle e^{ T k O } \rangle 
  \opsimeq_{T \to +\infty} 
\left[ \prod_{i=1}^N \int_0^1 d \rho_i^+ \right]   
e^{ \displaystyle  - T L_{2}^{[k]}  [  \rho_.^+   ]  } \opsimeq_{T \to +\infty} e^{ \displaystyle  T G(k)  }
\label{geneO2tasep}
\end{eqnarray}
So the scaled cumulants generation function $G(k)$ of Eq. \ref{gkper}
corresponds to the optimization of $  \left( -  L_{2}^{[k]}  [  \rho_.^+   ] \right) $ 
over the $N$ values of the empirical density $\rho_.^+ $ for $i=1,..,N$.

If one is interested only in the two first cumulants, one can use the analysis of the previous subsection
\ref{subsec_typfluctTASEP} as follows.
As recalled in Appendix \ref{app_reminder2.5}, the first cumulant $G_1$ of Eq. \ref{g1} corresponds to the steady state value $O_{st}$
involving the steady state $P_.^+$and the steady current $J$
\begin{eqnarray}
G_1 = \langle O_T \rangle = \alpha_0 +  \sum_{i=1}^N  \alpha_i P_i^{+}  +\nu J
\label{g1steadytasep}
\end{eqnarray}
The small typical fluctuations of order $\frac{1}{\sqrt{T} } $ around this steady state value 
can be rewritten in terms of the rescaled empirical observables 
$[{\hat \rho}_.^+ ; {\hat j}  ]$ of Eq. \ref{hatrhoqtasep}
\begin{eqnarray}
O_T - \langle O_T \rangle
 = \frac{1}{\sqrt{T} }   \left( 
   \sum_{i=1}^N  \alpha_i  {\hat \rho}^{+}_i  +\nu {\hat j} \right)
\label{additivesmallflucttasep}
\end{eqnarray}
so that the rescaled variance of Eq. \ref{g2} 
\begin{eqnarray}
G_2  \equiv  T   \langle  \left( O_T  - \langle  O_T  \rangle \right)^2  \rangle
= \bigg\langle \left( \sum_{i=1}^N  \alpha_i  {\hat \rho}^{+}_i  +\nu {\hat j}  \right)^2  \bigg\rangle
\label{g2evaltasep}
\end{eqnarray}
can be evaluated via the average over the Gaussian probability ${\hat P}^{[2.5]}_T [{\hat \rho}_.^+ ; {\hat j}  ] $ of the rescaled fluctuations of Eq. \ref{level2.5gausstasep}.


\section{ Conclusions  }

\label{sec_conclusion}

For a given inhomogeneous exclusion processes of $N$ sites between two reservoirs,
we have described how the trajectories probabilities allow to identify the relevant local empirical observables
and to obtain the corresponding rate function at Level 2.5. 
Since the only closure problem arises with the stationarity constraints,
we have considered the simplest approximation to close the hierarchy of the empirical dynamics, 
namely the Mean-Field approximation
for the empirical density of two consecutive sites, in direct correspondence with
the previously studied Mean-Field approximation for the steady state.
For a given inhomogeneous Totally Asymmetric model (TASEP), this Mean-Field approximation yields the large deviations for the joint distribution of the empirical density profile and of the empirical current, while the explicit contraction over the current allows to write the large deviations of the empirical density profile alone. For a given inhomogeneous Asymmetric model (ASEP), the local empirical observables also involve the empirical activities of the links and of the reservoirs, but the explicit contraction over these activities allows to write the large deviations for the joint distribution of the empirical density profile and of the empirical current. Finally, we have discussed the consequences for the large deviations properties of time-additive space-local observables.

To test the validity of the Mean-Field approximation for the large deviations properties of local empirical observables,
it would be very interesting to compare with numerical studies
 for various types of inhomogeneous samples, both for ASEP and for TASEP,
but this is left to future works since this clearly goes beyond the scope of the present article.


\appendix


\section{ Reminder on the large deviations at Level 2.5 for continuous-time Markov jump processes  }

\label{app_reminder2.5}

In this Appendix, we recall the dynamical large deviations that can be written  
for any continuous-time Markov jump process described by the Master Equation
\begin{eqnarray}
\frac{\partial P_t(C)}{\partial t} =    \sum_{C' }   W(C,C')  P_t(C') 
\label{mastereq}
\end{eqnarray}
for the probability $P_t(C)$ to be in configuration $C$ at time $t$.
The off-diagonal matrix elements $W(C,C') \geq 0$ represent the transitions rates from $C'$ to $C \ne C' $
while the diagonal elements are fixed by the conservation of probability to be
\begin{eqnarray}
W(C,C)   =  - \sum_{C' \ne C} W(C',C)
\label{wdiag}
\end{eqnarray}

The probability of the trajectory $[C(0 \leq t \leq T)] $ during the time-window $0 \leq t \leq T$
reads
\begin{eqnarray}
{\cal P}[C(0 \leq t \leq T)]   
=   e^{ \displaystyle    \sum_{t \in [0,T]: C(t^+) \ne C(t) } \ln (W(C(t^+),C(t))  ) +  \int_0^T dt W(C(t),C(t))   }
\label{pwtrajjump}
\end{eqnarray}

\subsection{ Large deviations at Level 2.5 for the empirical density and flows  }

For a trajectory $C(0 \leq t \leq T)$ over the large time-window $T$,
one focuses on the empirical time-averaged density
\begin{eqnarray}
 \rho(C) && \equiv \frac{1}{T} \int_0^T dt \  \delta_{C(t),C}  
 \label{rhoc}
\end{eqnarray}
satisfying the normalization
\begin{eqnarray}
\sum_C \rho(C) && = 1
\label{rhocnorma}
\end{eqnarray}
and on the jump density from $C$ to $C' \ne C$
\begin{eqnarray}
q(C',C) \equiv  \frac{1}{T} \sum_{t : C(t) \ne C(t^+)} \delta_{C(t^+),C'} \delta_{C(t),C} 
\label{jumpempiricaldensity}
\end{eqnarray}
satisfying the stationarity constraints (for any configurations $C$, the total incoming flow into $C$ should be equal to the total outgoing flow from $C$)
\begin{eqnarray}
\sum_{C' \ne C} q(C,C') = \sum_{C' \ne C} q(C',C)
\label{contrainteq}
\end{eqnarray}
The joint probability distribution of the empirical density $\rho(.)$ and flows $q(.,.)$
satisfy the following large deviation form with respect to the large time-window $T$ 
\cite{fortelle_thesis,fortelle_jump,maes_canonical,maes_onandbeyond,wynants_thesis,chetrite_formal,BFG1,BFG2,chetrite_HDR,c_ring,c_interactions,c_open,c_detailed,barato_periodic,chetrite_periodic,c_reset,c_inference,c_runandtumble,c_jumpdiff,c_skew,c_metastable,c_east}
\begin{eqnarray}
P^{[2.5]}_{T}[ \rho(.) ; q(.,.) ] \oppropto_{T \to +\infty} {\cal C}[ \rho(.) ; q(.,.) ] e^{- T I_{2.5}[ \rho(.) ; q(.,.) ] }
\label{level2.5master}
\end{eqnarray}
with the constitutive constraints already discussed in Eqs \ref{rhocnorma} and \ref{contrainteq}
\begin{eqnarray}
 {\cal C} [ \rho(.) ; q(.,.) ]
  = \delta \left( \sum_C \rho(C) - 1 \right) 
 \prod_C \left[  \sum_{C' \ne C} (q(C,C') - q(C',C) ) \right]
\label{constraints2.5master}
\end{eqnarray}
while the explicit rate function 
\begin{eqnarray}
I_{2.5}[ \rho(.) ; q(.,.) ]=  \sum_{C } \sum_{C' \ne C} 
\left[ q(C',C)  \ln \left( \frac{ q(C',C)  }{  W(C',C)  \rho(C) }  \right) 
 - q(C',C)  + W(C',C)  \rho(C)  \right]
\label{rate2.5master}
\end{eqnarray}
contains the relative entropy cost of having empirical flows $q(C',C) $
different from the typical flows $ W(C',C)  \rho(C) $
 that would be produced by the empirical density $   \rho(C) $.

The only way to satisfy the constitutive constraints of Eq. \ref{constraints2.5master}
and to make the rate function of Eq. \ref{rate2.5master} vanish
is when the empirical density $\rho(C) $ coincides with the steady state $P(C)$ 
of the master Equation \ref{mastereq} and when the empirical flows $ q(C',C) $
coincide with the steady state flows
\begin{eqnarray}
Q(C',C)  \equiv W(C',C)  P(C) 
\label{steadyflows}
\end{eqnarray}

\subsection{ Application to the large deviations properties of time-additive observables  }

The empirical density $\rho(C)$ of Eq. \ref{rhoc} and the empirical flows $q(C',C) $ of Eq. \ref{jumpempiricaldensity}
allow to reconstruct any time-additive observable $O_T$
via the introduction of appropriate coefficients $[\alpha(C) ;\beta (C',C)]$
\begin{eqnarray}
O_T  = \sum_C \left[ \alpha(C)  \rho(C) + \sum_{C' \ne C} \beta (C',C) q(C',C)\right]
 = \frac{1}{T} \int_0^T dt \  \alpha(C(t))  + \frac{1}{T} \sum_{t : C(t) \ne C(t^+)} \beta(C(t^+),C(t))
 \label{additive}
\end{eqnarray}

The large deviations properties of this observable for large $T$
 \begin{eqnarray}
 P_T( O ) \opsimeq_{T \to +\infty} e^{- T I ( O )}
\label{level1def}
\end{eqnarray} 
are governed by the rate function $I(O) \geq 0$ that vanishes only for the steady state value $O_{st}$
 \begin{eqnarray}
 I ( O_{st} ) =0
\label{iaeqvanish}
\end{eqnarray}
that can be reconstructed via Eq. \ref{additive}
from the steady state $P(C)$ and from the steady state flows $Q(C',C) $ of Eq. \ref{steadyflows}
\begin{eqnarray}
O_{st} && = \sum_C \left[ \alpha(C)  P(C) + \sum_{C' \ne C} \beta (C',C) Q(C',C)\right]
\label{additivesteady}
\end{eqnarray}

Equivalently, one can focus on the generation function $G(k)$ of the scaled cumulants $G_n$
\begin{eqnarray}
G(k)  = \sum_{n=1}^{+\infty} G_n \frac{k^n}{n!}   = G_1 k + G_2 \frac{k^2}{2}  + O(k^3)
\label{gkper}
\end{eqnarray}
where the averaged value $G_1$ corresponds to the steady state value of Eq. \ref{additivesteady}
\begin{eqnarray}
G_1 = \langle O_T \rangle = O_{st} 
\label{g1}
\end{eqnarray}
while $G_2$ corresponds to the rescaled variance
\begin{eqnarray}
G_2  \equiv  T   \langle  \left( O_T  - \langle  O_T  \rangle \right)^2  \rangle
\label{g2}
\end{eqnarray}

The scaled cumulant generation function $G(k)$ of Eq. \ref{gkper}
governs the large $T$ behavior of the generating function 
\begin{eqnarray}
\langle e^{ T k O } \rangle \equiv \int_{-\infty}^{+\infty} dO P_T ( O )e^{ T k O} 
\opsimeq_{T \to +\infty}  \int_{-\infty}^{+\infty} dO 
e^{ \displaystyle  T \left( - I (O)+ k O \right) }
\opsimeq_{T \to +\infty} e^{ \displaystyle  T G(k)  }
\label{geneOdef}
\end{eqnarray}
The saddle-point evaluation of the above integral over $O$ above
 yields that the scaled cumulant generation function $G(k)$ corresponds to the Legendre transform of the rate function $I(O)$
\begin{eqnarray}
 - I (O)+ k O && = G(k)  
 \nonumber \\
 - I'(O) + k && =0
\label{legendre}
\end{eqnarray}

The generating function of the additive observable of Eq. \ref{additive}
can be evaluated from the joint probability $P^{[2.5]}_T [ \rho(.) ; q(.,.)  ] $ at Level 2.5 of Eq. \ref{level2.5master}
\begin{eqnarray}
&& \langle e^{ T k O } \rangle 
 =  \int d \rho(.) \int d q(.,.) P^{[2.5]}_{T}[ \rho(.) ; q(.,.) ] 
 e^{ \displaystyle T k  \sum_C \left( \alpha(C)  \rho(C) + \sum_{C' \ne C} \beta (C',C) q(C',C)\right) }
\nonumber \\
&& \oppropto_{T \to +\infty}  \int d \rho(.) \int d q(.,.)
{\cal C}[ \rho(.) ; q(.,.) ] 
e^{\displaystyle T \left[ - I_{2.5}[ \rho(.) ; q(.,.) ]  +k  \sum_C \left( \alpha(C)  \rho(C) + \sum_{C' \ne C} \beta (C',C) q(C',C)\right) \right]}
\label{geneO}
\end{eqnarray}
So the scaled cumulant generation function $G(k)$ can be obtained via the saddle-point evaluation of this integral over empirical observables $[ \rho(.) ; q(.,.) ]  $ respecting the constitutive constraints ${\cal C}[ \rho(.) ; q(.,.) ]  $.


\section{ Large deviations in the whole configuration space of inhomogeneous exclusion models }

\label{app_config}

The general framework recalled in Appendix \ref{app_reminder2.5}
can be applied to inhomogeneous exclusion models as follows.

\subsection{ Application of the large deviations at Level 2.5 in the whole configuration space  }

For a trajectory $\{ S_1(t),...,S_N(t)\}$ of the $N$ spins over the large time-window $0 \leq t \leq T$,
the empirical time-averaged density of Eq. \ref{rhoc}
\begin{eqnarray}
 \rho(S_1,...,S_N) && \equiv \frac{1}{T} \int_0^T dt \  \prod_{n=1}^N \delta_{S_n(t),S_n}  
 \label{rho}
\end{eqnarray}
satisfies the normalization of Eq. \ref{rhocnorma}
\begin{eqnarray}
\sum_{S_1=\pm} ... \sum_{S_N=\pm}  \rho(S_1,...,S_N) && = 1
\label{rho1norma}
\end{eqnarray}
while the empirical flows of Eq. \ref{jumpempiricaldensity} associated to the flip rates of the model are 
defined as follows.

(i) For each bulk link $(i+1/2)$ with $i=1,...,N-1$, the two rates $w_{i+1/2}^{\pm}$ will produce the empirical flows
\begin{eqnarray}
q_{i+1/2}^{+} (S_1,..,S_{i-1} ; S_{i+2} ,..,S_N)
\equiv  \frac{1}{T}    \sum_{t \in [0,T] : \substack{ S_i(t)=+ ;  S_{i+1}(t)=-   \\ S_i(t^+)= - ; S_{i+1}(t^+)=+}} 
\left[ \prod_{n=1}^{i-1} \delta_{S_n(t),S_n}  \right] \left[ \prod_{p=i+2}^{N} \delta_{S_p(t),S_p}  \right] 
\nonumber \\
q_{i+1/2}^{-} (S_1,..,S_{i-1} ; S_{i+2} ,..,S_N)
\equiv  \frac{1}{T}    \sum_{t \in [0,T] : \substack{ S_i(t)=- ;  S_{i+1}(t)=+   \\ S_i(t^+)= + ; S_{i+1}(t^+)=-}} 
\left[ \prod_{n=1}^{i-1} \delta_{S_n(t),S_n}  \right] \left[ \prod_{p=i+2}^{N} \delta_{S_p(t),S_p}  \right] 
\label{qbulk}
\end{eqnarray}

(ii) For the boundary spin $S_1$ in contact with the Left reservoir, the two rates $w_1^{\pm}$ will produce the empirical flows
\begin{eqnarray}
q_1^{+} (S_2,...,S_N) \equiv  \frac{1}{T}    \sum_{t \in [0,T] :  \substack{S_1(t)=+ \\   S_1(t^+)= - }} 
 \left[ \prod_{n=2}^{N} \delta_{S_n(t),S_n}  \right] 
\nonumber \\
q_1^{-} (S_2,...,S_N) \equiv  \frac{1}{T}    \sum_{t \in [0,T] :  \substack{S_1(t)=- \\   S_1(t^+)= + }} 
 \left[ \prod_{n=2}^{N} \delta_{S_n(t),S_n}  \right] 
\label{qleft}
\end{eqnarray}
while for the boundary spin $S_N$ in contact with the Right reservoir,
the two rates $w_N^{\pm}$ will produce the empirical flows
\begin{eqnarray}
q_N^{+} (S_1,...,S_{N-1}) \equiv  \frac{1}{T}    \sum_{t \in [0,T] :  \substack{S_N(t)=+ \\  S_N(t^+)= - }} 
 \left[ \prod_{n=1}^{N-1} \delta_{S_n(t),S_n}  \right] 
\nonumber \\
q_N^{-} (S_1,...,S_{N-1}) \equiv  \frac{1}{T}    \sum_{t \in [0,T] :  \substack{S_N(t)=- \\   S_N(t^+)= + }} 
 \left[ \prod_{n=1}^{N-1} \delta_{S_n(t),S_n}  \right] 
\label{qright}
\end{eqnarray}

The stationarity constraint of Eq. \ref{contrainteq} for the empirical density $\rho(S_1,...,S_N)$ reads
\begin{eqnarray}
&& 0  = \partial_t \rho(S_1,...,S_N)
\label{contrainteqs}
\nonumber \\
&&=  \sum_{i=1 }^{N-1} 
 \left( \delta_{S_i,-} \delta_{S_{i+1},+} -  \delta_{S_i,+} \delta_{S_{i+1},-}  \right)
 \left[ q_{i+1/2}^{+}  (S_1,..,S_{i-1} ; S_{i+2} ,..,S_N)
 - q_{i+1/2}^{-}  (S_1,..,S_{i-1} ; S_{i+2} ,..,S_N)  \right]
 \nonumber \\
 && + \left( \delta_{S_1,-}  -  \delta_{S_1,+}   \right)
 \left[  q_1^{+} (S_2,...,S_N)  - q_1^{-} (S_2,...,S_N)   \right]
 + \left( \delta_{S_N,-}  -  \delta_{S_N,+}   \right)
 \left[ q_N^{+} (S_1,...,S_{N-1}) - q_N^{-} (S_1,...,S_{N-1})   \right]
\nonumber
\end{eqnarray}

The rate function of Eq. \ref{rate2.5master}
that governs the large deviations at Level 2.5 of Eq. \ref{level2.5master} 
reads for the present model
\begin{footnotesize}
\begin{eqnarray}
&& I_{2.5}[ \rho(.) ; q_.^{\pm}(.) ]
 =  \sum_{S_1 = \pm}  \sum_{S_2 = \pm} ... \sum_{S_N = \pm}
\nonumber \\
&& \bigg( \sum_{i=1}^{N-1} \delta_{S_i,+} \delta_{S_{i+1},-}
\bigg[ q_{i+1/2}^{+} (..,S_{i-1} ; S_{i+2} ,..)  \ln \left( \frac{q_{i+1/2}^{+} (..,S_{i-1} ; S_{i+2} ,..) }
{  w_{i+1/2}^{+}   \rho(..S_{i-1},+,-,S_{i+2}..) }  \right) 
- q_{i+1/2}^{+} (..,S_{i-1} ; S_{i+2} ,..)  + w_{i+1/2}^{+}  \rho(..S_{i-1},+,-,S_{i+2}..)  \bigg]
\nonumber \\
&& +\sum_{i=1}^{N-1} \delta_{S_i,-} \delta_{S_{i+1},+}
\bigg[ q_{i+1/2}^{-} (..,S_{i-1} ; S_{i+2} ,..)  \ln \left( \frac{q_{i+1/2}^{-} (..,S_{i-1} ; S_{i+2} ,..) }
{  w_{i+1/2}^{-}   \rho(..S_{i-1},-,+,S_{i+2}..) }  \right) 
 - q_{i+1/2}^{-} (..,S_{i-1} ; S_{i+2} ,..)  + w_{i+1/2}^{-}  \rho(..S_{i-1},-,+,S_{i+2}..)  \bigg]
\nonumber \\
&& +  
\left[ q_1^{S_1} (S_2,..) \ln \left( \frac{q_1^{S_1} (S_2,..)  }{  w_1^{S_1}  \rho(S_1,S_2,..) } \right) 
 - q_1^{S_1} (S_2,..)  + w_1^{S_1}  \rho(S_1,S_2,..)  \right]
 \nonumber \\ &&
 +  \left[ q_N^{S_N} (..,S_{N-1}) \ln \left( \frac{q_N^{S_N} (..,S_{N-1})  }{  w_N^{S_N}  \rho(..,S_{N-1},S_N) } \right) 
 - q_N^{S_N} (..,S_{N-1})  + w_N^{S_N}  \rho(..,S_{N-1},S_N)  \right]
\label{rate2.5}
\end{eqnarray}
\end{footnotesize}


\subsection{ Application to the large deviation properties of time-additive observables  }

For the present model, the most general additive observable of Eq. \ref{additive}
involves coefficients $[ \alpha(.) ; \beta_.^{\pm}(.) ] $ associated to 
the empirical observables $[ \rho(.) ; q_.^{\pm}(.) ]$ in the whole configuration space
\begin{eqnarray}
 O_T 
&& =  \left[ \prod_{k=1}^{N} \sum_{S_k=\pm}  \right] 
\left(  \alpha(S_1,...,S_N) \rho(S_1,...,S_N) + \beta_1^{S_1} (S_2,..,S_N) q_1^{S_1} (S_2,..,S_N)
+ \beta_N^{S_N} (S_1,..,S_{N-1}) q_N^{S_N} (S_1,..,S_{N-1}) \right)
\nonumber \\
&& + \sum_{i=1}^{N-1} 
\left[ \prod_{n=1}^{i-1} \sum_{S_n=\pm}  \right] \left[ \prod_{p=i+2}^{N}\sum_{S_p=\pm}   \right]
\sum_{\epsilon=\pm} \beta_{i+1/2}^{\epsilon} (..,S_{i-1} ; S_{i+2} ,..) q_{i+1/2}^{\epsilon} (..,S_{i-1} ; S_{i+2} ,..)
\label{additiveglobal}
\end{eqnarray}


\subsection{ Discussion  }

The empirical density $\rho(S_1,...,S_N) $ 
and the empirical flows $ q_.^{\pm}(.) $ described above
have been defined in the space of the $2^N$ configurations of the $N$ spins,
while one would like to analyze instead the local empirical observables involving only 
one spin or two consecutive spins.
Similarly, the general time-additive observable of Eq. \ref{additiveglobal}
involve coefficients $[ \alpha(.) ; \beta_.^{\pm}(.) ] $ depending on the whole configuration,
while one is often more interested into time-additive observables 
that are made of contributions that are local in space.
More generally, whenever the dynamical rules of a many-body model are local in space,
the large deviations at Level 2.5 in the full configuration space are somewhat 'overkill',
and it is then natural to try to analyze the dynamics via the appropriate local empirical observables,
as discussed in more details in the Introduction to motivate the approach described in the main text.



\begin{thebibliography}{99}



\bibitem{oono}
Y. Oono,
Progress of Theoretical Physics Supplement 99, 165 (1989).

\bibitem{ellis}
R.S. Ellis, Physica D 133, 106 (1999).

\bibitem{review_touchette}
H. Touchette, Phys. Rep. 478, 1 (2009).


\bibitem{fortelle_thesis}
A. de La Fortelle, PhD (2000)
"Contributions to the theory of large deviations and applications" INRIA Rocquencourt.

\bibitem{fortelle_chain}
G. Fayolle and A. de La Fortelle,
Problems of Information Transmission 38, 354 (2002).


\bibitem{c_largedevdisorder}
C. Monthus, Eur. Phys. J. B 92, 149 (2019) in the
topical issue " Recent Advances in the Theory of Disordered Systems"
edited by F. Igloi and H. Rieger.

\bibitem{c_reset}
C. Monthus, J. Stat. Mech. (2021) 033201.


\bibitem{c_inference}
C. Monthus,  J. Stat. Mech. (2021) 063211. 



\bibitem{fortelle_jump}
A. de La Fortelle, 
Problems of Information Transmission 37 , 120 (2001).



\bibitem{maes_canonical}
C. Maes and K. Netocny, Europhys. Lett. 82, 30003 (2008).

\bibitem{maes_onandbeyond}
C. Maes, K. Netocny and B. Wynants, Markov Proc. Rel. Fields. 14, 445 (2008).

\bibitem{wynants_thesis}
B. Wynants, arXiv:1011.4210, PhD Thesis (2010), "Structures of Nonequilibrium Fluctuations", Catholic University of Leuven.


\bibitem{chetrite_formal}
A. C. Barato and R. Ch\'etrite, J. Stat. Phys. 160, 1154 (2015).

\bibitem{BFG1}
L. Bertini, A. Faggionato and D. Gabrielli, 
Ann. Inst. Henri Poincare Prob. and Stat. 51, 867 (2015).

\bibitem{BFG2}
L. Bertini, A. Faggionato and D. Gabrielli, 
Stoch. Process. Appli. 125, 2786 (2015).

\bibitem{chetrite_HDR}
R. Ch\'etrite, HDR Thesis (2018)
"P\'er\'egrinations sur les ph\'enom\`enes al\'eatoires dans la nature",
 Laboratoire J.A. Dieudonn\'e, Universit\' e de Nice.

\bibitem{c_ring}
C. Monthus, J. Stat. Mech. (2019) 023206.

\bibitem{c_interactions}
C. Monthus, J. Phys. A: Math. Theor. 52, 135003 (2019).


\bibitem{c_open}
C. Monthus, J. Phys. A: Math. Theor. 52, 025001 (2019).

\bibitem{c_detailed}
C. Monthus, J. Phys. A: Math. Theor. 52, 485001 (2019).

\bibitem{barato_periodic}
A. C. Barato, R. Ch\'etrite, J. Stat. Mech. (2018) 053207.

\bibitem{chetrite_periodic}
L. Chabane, R. Ch\'etrite, G. Verley, J. Stat. Mech. (2020) 033208.

 \bibitem{c_runandtumble}
C. Monthus, J. Stat. Mech. (2021) 083212.

\bibitem{c_jumpdiff}
C. Monthus,  J. Stat. Mech. (2021) 083205.

\bibitem{c_skew}
C. Monthus,  arXiv:2106.09429

\bibitem{c_metastable}
C. Monthus,  arXiv:2107.05354.

\bibitem{c_east}
C. Monthus,  arXiv:2109.05924.




\bibitem{maes_diffusion}
C. Maes, K. Netocny and B.  Wynants
Physica A 387, 2675 (2008).


\bibitem{engel}
J. Hoppenau, D. Nickelsen and A. Engel,
 New J. Phys. 18 083010 (2016).

\bibitem{c_lyapunov}
C. Monthus, J. Stat. Mech. (2021) 033303.

\bibitem{coghi}
F. Coghi, R. Ch\'etrite and H. Touchette, Phys. Rev. E 103, 062142 (2021).



\bibitem{derrida-lecture}
B. Derrida, J. Stat. Mech. P07023 (2007).

\bibitem{sollich_review}
R. L. Jack, P. Sollich, The European Physical Journal Special Topics  224, 2351 (2015).

\bibitem{lazarescu_companion}
A. Lazarescu, J. Phys. A: Math. Theor. 48 503001 (2015).

\bibitem{lazarescu_generic}
A. Lazarescu, J. Phys. A: Math. Theor. 50 254004 (2017).

\bibitem{jack_review}
R. L. Jack, Eur. Phy. J. B  93, 74 (2020).

\bibitem{vivien_thesis}
V. Lecomte, PhD Thesis (2007)
"Thermodynamique des histoires et fluctuations hors d'\'equilibre"
Universit\'e Paris.



\bibitem{lecomte_chaotic}
V. Lecomte, C. Appert-Rolland and F. van Wijland,
Phys. Rev. Lett. 95, 010601 (2005).

\bibitem{lecomte_thermo}
V. Lecomte, C. Appert-Rolland and F. van Wijland,
J. Stat. Phys. 127, 51 (2007).

\bibitem{lecomte_formalism}
V. Lecomte, C. Appert-Rolland and F. van Wijland,
Comptes Rendus Physique 8, 609 (2007).

\bibitem{lecomte_glass}
J.P. Garrahan, R.L. Jack, V. Lecomte, E. Pitard, K. van Duijvendijk, F. van Wijland,
Phys. Rev. Lett. 98, 195702 (2007).

\bibitem{kristina1}
J.P. Garrahan, R.L. Jack, V. Lecomte, E. Pitard, K. van Duijvendijk and F. van Wijland, 
J. Phys. A 42, 075007 (2009).

\bibitem{kristina2}
K. van Duijvendijk, R.L. Jack and F. van Wijland, 
Phys. Rev. E 81, 011110 (2010).


\bibitem{jack_ensemble}
R. L. Jack, P. Sollich, Prog. Theor. Phys. Supp. 184, 304 (2010).

\bibitem{simon1}
D. Simon, J. Stat. Mech. (2009) P07017.



\bibitem{simon2}
V. Popkov, G. M. Schuetz, D. Simon, J. Stat. Mech. P10007 (2010).

\bibitem{simon3}
D. Simon, J. Stat. Phys. 142,  931 (2011).

\bibitem{Gunter1}
V. Popkov, G. M. Schuetz, J. Stat. Phys 142,  627 (2011)


\bibitem{Gunter2}
V. Belitsky, G. M. Schuetz, J. Stat. Phys. 152, 93 (2013).


\bibitem{Gunter3}
O. Hirschberg, D. Mukamel, G. M. Schuetz, J. Stat. Mech. P11023 (2015).

\bibitem{Gunter4}
G. M. Schuetz, From Particle Systems to Partial Differential Equations II, Springer Proceedings in Mathematics and Statistics Volume 129, pp 371-393, P. Goncalves and A.J. Soares (Eds.), (Springer, Cham, 2015).





\bibitem{chetrite_canonical}
R. Ch\'etrite and H. Touchette,
Phys. Rev. Lett. 111, 120601 (2013).

\bibitem{chetrite_conditioned}
R. Ch\'etrite and H. Touchette
 Ann. Henri Poincare 16, 2005 (2015).

\bibitem{chetrite_optimal}
R. Ch\'etrite, H. Touchette, J. Stat. Mech. P12001 (2015).



\bibitem{touchette_circle}
P. T. Nyawo, H. Touchette, Phys. Rev. E 94, 032101 (2016).

\bibitem{touchette_langevin}
H. Touchette, Physica A 504, 5 (2018).

\bibitem{touchette_occ}
F. Angeletti, H. Touchette, Journal of Mathematical Physics 57, 023303 (2016).

\bibitem{touchette_occupation}
P. T. Nyawo, H. Touchette, Europhys. Lett. 116, 50009 (2016); \\
P. T. Nyawo, H. Touchette, Phys. Rev. E 98, 052103 (2018).

\bibitem{derrida-conditioned}
B. Derrida and T. Sadhu, Journal of Statistical Physics 176, 773 (2019); \\
B. Derrida and T. Sadhu, 
Journal of Statistical Physics 177, 151 (2019).

\bibitem{derrida-ring}
K. Proesmans, B. Derrida, J. Stat. Mech. (2019) 023201.

\bibitem{bertin-conditioned}
N. Tizon-Escamilla, V. Lecomte and E. Bertin, 	J. Stat. Mech. (2019) 013201.



\bibitem{touchette-reflected}
J. du Buisson, H. Touchette, Phys. Rev. E 102, 012148 (2020).

\bibitem{touchette-reflectedbis} 
E. Mallmin, J. du Buisson and H. Touchette,  J. Phys. A: Math. Theor. 54 295001 (2021).


\bibitem{previousquantum2.5doob}
F. Carollo, J. P. Garrahan, I. Lesanovsky, C. Perez-Espigares, Phys. Rev. A 98, 010103 (2018).

 \bibitem{quantum2.5doob}
F. Carollo, R. L. Jack, J. P. Garrahan, Phys. Rev. Lett. 122, 130605 (2019).

 \bibitem{quantum2.5dooblong} 
F. Carollo, J. P. Garrahan, R. L. Jack, J. Stat. Phys. 184, 13 (2021).


\bibitem{c_ruelle}
C. Monthus,  J. Stat. Mech. (2021) 063301.


\bibitem{lapolla}
A. Lapolla, D. Hartich, A. Godec, Phys. Rev. Research 2, 043084 (2020).


\bibitem{mft}
L. Bertini, A. De Sole, D. Gabrielli, G. Jona-Lasinio, and C. Landim
Rev. Mod. Phys. 87, 593 (2015).


\bibitem{chemical}
 A. Lazarescu, T. Cossetto, G. Falasco and  M. Esposito,
J. Chem. Phys. 151, 064117 (2019)


\bibitem{chabane}
L. Chabane, A. Lazarescu and G. Verley, arXiv:2109.06830 




\bibitem{1defect}
S.A. Janowski and J.L. Lebowitz, Phys. Rev. A45, 618 (1992).

\bibitem{greulich_defects}
P. Greulich and A. Schadschneider, Physica A 387, 1972 (2008).

\bibitem{2defects}
Niladri Sarkar and Abhik Basu, Phys. Rev. E 90, 022109 (2014).


\bibitem{smooth}
R. B. Stinchcombe and S. L. A. de Queiroz, Phys. Rev. E 83, 061113 (2011).

\bibitem{inhomo}
A. Goswami, M. Chatterjee, S. Mukherjee, arXiv:2107.10185.




\bibitem{stanley}
E. Koscielny-Bunde, A. Bunde, S. Havlin and H. E. Stanley,
Phys. Rev. A 37, 1821(1988).

\bibitem{barma_short}
G. Tripathy and M. Barma,
Phys. Rev. Lett. 78, 3039 (1997).

\bibitem{barma_long}
G. Tripathy and M. Barma,
Phys. Rev. E 58, 1911 (1998).

\bibitem{goldstein}
S. Goldstein and E. R. Speer,
Phys. Rev. E 58, 4226 (1998).

\bibitem{krug}
J. Krug, Braz. J. Phys. 30, 97 (2000).

\bibitem{DTASEP}
K. M. Kolwankar and A. Punnoose,
Phys. Rev. E 61, 2453 (2000).



\bibitem{enaud}
C. Enaud and B. Derrida, Europhys. Lett. 66, 83 (2004).


\bibitem{MFtasepHarris}
R. J. Harris and R. B. Stinchcombe
Phys. Rev. E 70, 016108 (2004).

\bibitem{rgshort}
R. Juhasz, L. Santen and F. Igloi, Phys. Rev. Lett. 94, 010601 (2005).

\bibitem{rglong}
R. Juhasz, L. Santen and F. Igloi, Phys. Rev. E74, 061101 (2006).

\bibitem{rgextreme}
R. Juhasz, Y.-C. Lin, and F. Igloi
Phys. Rev. B 73, 224206 (2006).

\bibitem{barma}
M. Barma, Physica A 372, 22  (2006).

\bibitem{MFtasep}
M. E. Foulaadvand, S. Chaaboki and M. Saalehi, Phys. Rev. E 75, 011127 (2007).



\bibitem{greulich}
P. Greulich and A. Schadschneider, J. Stat. Mech. (2008) P04009.



\bibitem{MFasep}
R. Juhasz, J. Stat. Mech. (2011) P11010.

\bibitem{revisited}
J. Szavits-Nossan, J. Phys. A: Math. Theor. 46 (2013) 315001.

\bibitem{india}
T. Banerjee, N. Sarkar and A. Basu, J. Stat. Mech. (2015) P01024.

\bibitem{india2020}
A. Haldar and A. Basu,
Phys. Rev. Research 2, 043073 (2020).

\bibitem{indian}
T. Banerjee and A. Basu, Phys. Rev. Research 2, 013025 (2020).

\bibitem{bayesian}
M. Cavallaro, Y. Wang, D. Hebenstreit, R. Dutta, arXiv:2109.05100


\bibitem{c_lindbladExclusion}
C. Monthus, J. Stat. Mech. (2017) 043302.




\bibitem{MFpure}
B. Derrida, E. Domany and D. Mukamel,
Journal of Statistical Physics 69, 667 (1992).








\bibitem{galla}
E. G. D. Cohen and G. Gallavotti, Journal of Statistical Physics, 96, 1343 (1999)

\bibitem{kurchan_langevin}
J. Kurchan, J. Phys. A: Math. Gen. 31 3719 (1998).

\bibitem{Leb_spo}
J.L. Lebowitz and H. Spohn, J. Stat. Phys. 95, 333 (1999).

\bibitem{maes1999}
C. Maes, J. Stat. Phys. 95, 367 (1999).

\bibitem{jepps}
O. Jepps, D. J. Evans, D. J. Searles, Physica D, 187, 326 (2004).

\bibitem{harris_Schu}
R. J. Harris and G. M. Sch\"utz,
J. Stat. Mech.  P07020 (2007).


\bibitem{kurchan}
J. Kurchan J. Stat. Mech. (2007) P07005.

\bibitem{searles}
E.M. Sevick, R. Prabhakar, S. R. Williams, D. J. Searles,
Ann. Rev. of Phys. Chem.  Vol 59, 603 (2008). 

\bibitem{zia}
R. K. P. Zia and B. Schmittmann J. Stat. Mech. P07012 (2007).

\bibitem{chetrite_thesis}
R. Ch\'etrite, PhD Thesis 2008 
"Grandes d\'eviations et relations de fluctuation dans certains mod\`eles de syst\`emes
hors d'\'equilibre"  ENS Lyon.


\bibitem{maes2009}
C. Maes, K. Netocny, and B. Shergelashvili, {\it A selection of 
nonequilibrium issues}, In Methods of Contemporary Mathematical Statistical 
Physics, R. Koteck\'y ed. Lecture notes in Mathematics, Vol. 1970 (2009) 247.

\bibitem{maes2017}
C. Maes, J. Phys. A 50, 381001  (2017).




\end{thebibliography}
\end{document}